\documentclass[aps,prd,preprint]{revtex4}
\usepackage{CJK}

\begin{document}

\newcommand{\be}{\begin{equation}}
\newcommand{\ee}{\end{equation}}


\author{Alberto A. Garcia--Diaz }
\altaffiliation{aagarcia@fis.cinvestav.mx}  \author{Gustavo
Gutierrez--Cano}

\affiliation{Departmento~de~F\'{\i}sica,
~Centro~de~Investigaci\'on~y~de~Estudios~Avanzados~del~IPN,\\
 Av. IPN 2508, C.P. 07360, Col. San Pedro Zacatenco, Ciudad de M\'exico,
 M\'exico.\\}

 \title{Regularity conditions for spherically
 symmetric solutions of Einstein-nonlinear electrodynamics equations; revised and improved version}
 \date{\today}

\topmargin -2cm

 \vspace{0.5cm}

\begin{abstract}
In this report, the regularity conditions at the center for static
spherically symmetric (SSS) solutions of the Einstein equations
coupled to nonlinear electrodynamics (NLE) with Lagrangian
$\mathcal{L}= \mathcal{L}(\mathcal{F})$, depending on the
electromagnetic invariant $\mathcal{F}=F_{\mu\nu}\,F^{\mu\nu}/4$,
are established. The traceless Ricci (TR) tensor eigenvalue $S$, the
Weyl tensor eigenvalue $\Psi_2$ and the scalar curvature $R$
characterize the independent Riemman tensor invariants of SSS
metrics. The necessary and sufficient regularity conditions for
electric NLE SSS solutions require $\lim_{r\rightarrow
0}\{\Psi_2,S,R\}\rightarrow \{0,0,(0,4\Lambda+ 4\mathcal{L}(0))\}$,
such that the metric function $Q(r)$ and the electric field
$q_0F_{rt}=:\mathcal{E}$ behave as $\{Q,\dot Q,\ddot
Q\}\rightarrow\{0,0,2\}$ and
$\{\mathcal{E},\dot\mathcal{E},\ddot\mathcal{E}\}\rightarrow\{0,0,0\}$,
 as $r\rightarrow 0$. The general linear integral representation
of the electric NLE SSS metric in terms of an arbitrary electric
field $\mathcal{E}$, together with $\{\Psi_2,S,R\}$, is explicitly
given. Moreover, beside the regular or singular behavior at the center,
these solutions may exhibit different asymptotic behavior at spatial
infinity such as the Reissner--Nordtr\"om (Maxwell) asymptotic, or
present the dS--AdS or other kind of asymptotic.
\end{abstract}
\maketitle

\tableofcontents

\section{Introduction}

 In our previous publication \cite{Garcia-Diaz:2019acq}
dealing with a static spherically symmetric (SSS) metric in the
framework of the Einstein--nonlinear electrodynamics theory, it was
established that a linear superposition principle of solutions of
different kinds holds. In this report we focus mostly in the
determination of the conditions under which a SSS metric is regular
(singular--free) at the center (origin of coordinates).

Nowadays ``black hole'' has become a term of everyday use; in
cinema, television programs, and internet, there are shown scenes
about the evolution, collapse, interaction of black holes, emission
of jets of energy, using the resources of numerical mathematical
simulation and the abilities of the multimedia professionals. In
spite of the large theoretical progress, and astrophysical advances
in the black hole physics there still remain some incognita about
their final singular stage, if any, as predicted by the theory, or,
on the contrary, they exhibit a regular behavior allowing to avoid
the singular catastrophe. The determination of the regularity
conditions is the problem we posed in this report for a certain kind
of metrics, namely, the static spherically symmetric spacetimes.

The most famous black holes are described by the static spherically
symmetric (SSS) Schwarzschild solution \cite{Schwarzschild1916}, the
first exact solution derived in 1916  for a point mass $m$, and the
stationary axial symmetric Kerr solution \cite{Kerr}, 1963, for a
rotating mass. The charged point mass solution was reported by
Reissner \cite{Reissner}, 1916, and Nordstr\"om \cite{Nordstrom},
1918, while the charged rotating mass black hole solution was
reported by Newman and collaborators \cite{Newman:1965my}. Vacuum
and Maxwell charged black hole solutions are all singular at the
center.

Oppenheimer et al. \cite{OppenheimerSnyder39}, and
\cite{OppenheimerVolkoff39}, at the end of the 30's studied the
collapse of a massive spherically symmetric star and came to the
conclusion that in the process (time dependent) of approaching to
its critical surface, the star will increase indefinitely its
curvature, and that the light radiated by an imploding star will be
red--shifted, and, as it reaches its critical radius, the redshift
will become infinite and the star will disappear from the observer's
sight; it becomes black behind the event horizon. The static
spherically symmetric representation of the Schwarzschild metric
exhibits an ``apparent singularity''  or a ``Schwarzschild
singularity'' at the Schwarzschild radius $r=2m$; various attempt
were done, by means of new coordinates, to remove the ``apparent
singularity.'' A real success  was achieved  by the introduction  of
 null coordinates by Finkelstein \cite{Finkelstein:1958zz}, 1959,
Kruskal \cite{Kruskal:1959vx}, 1960, and Szekeres
\cite{Szekeres:1960gm}, 1960, for the maximal extension of the
Schwarzschild metric  and the interpretation of the Schwarzschild
radius as the surface bound to the event horizon. Later, in 1963,
Kerr reported his rotating black hole solution; this discovery
catapulted a period of intense activity in the field of black hole
theory. To get an insight into black holes Penrose introduced
spinors in the description of spacetimes, 1960 and 1972, see \cite
{Penrose86}, and developed the so called Penrose diagram procedure.
Hawking \cite{Hawking73} in the beginning of the 70's, formulated
the thermodynamics of black holes. Meanwhile, in differential
geometry various tools become of common use. Newmann and Penrose
introduced in general relativity their tetrad formalism. Petrov
\cite{Petrov66}, published the algebraic classification of the
Riemman and Weyl tensors which allows a deeper understanding of the
algebraic spacetime structures. Pleba\'nski \cite{Plebanski64},
1964, reported the classification of the energy--momentum tensor.

It was only until recently that regular black holes erupted in the
scene in a formal way supported by the nonlinear electrodynamics
(NLE).  Ay\'on--Beato and Garc\'\i a published the first regular
static spherically symmetric charged solution \cite{AyonGarcia:1998}
described in terms of NLE potentials; this contribution caused a big
impulse in the study of regular solutions in Einstein gravity; these
authors also found the nonlinear electrodynamics source (magnetic
potential) \cite{AyonGarciaOnBardeen} to the Bardeen model
\cite{Bardeen68} reported in 1968; the first metric structure
fitting the tensor energy conditions exhibiting curvature regularity
everywhere.

Born and Infeld (BI)~\cite{BornInfeld1934}, 1934, formulated
nonlinear electrodynamics to provide the electron with a spatial
environment supporting a charge with finite self--energy.
Pleba\'nski \cite{Plebanski70}, 1970, published a generalization of
the BI NLE. Later Garc{\'\i a, Salazar, and
Pleba\'nski~\cite{SalazarGarciaPleb2} reported all Petrov type D
solutions to Einstein--Born--Infeld electrodynamics allowing for
stationary and axial symmetries, among them the static spherically
symmetric solution; they also developed NLE allowing for duality
rotations~\cite{SalazarGarciaPleb1} and studied the birefringence
properties of the theory~\cite{IbarguenGP88}. Fradkin and Tseytlin
\cite{FradkinTseytlin85}, and Gibbons and Rasheed
\cite{GibbonsRasheed95} established that BI--NLE arises at the low
energy limit of the string theory; this fact leads to a renewed
interest on BI--NLE.  In the last decade various families of the SSS
solutions of Einstein--NLE equations have been reported (the number
of articles is large), and attempts to derive regular stationary
rotating solutions have been undertaken by various researchers
without notorious successes.

In Einstein theory to establish the regularity of solutions one
analyzes the behavior of the Riemann tensor invariants of the
gravitational field; it is well--known that, in general, a pseudo
Riemannian spacetime possesses 14 independent curvature invariants,
see Thomas \cite{Thomas34}, Weinberg \cite{Weinberg72}, and, for a
more recent and complete work, see Zakhary and McIntosh
\cite{McIntoshZakhary97}. Among these invariants one finds the
widely known and commonly evaluated Kretschmann quadratic Riemann
scalar $R_{\alpha\beta\gamma\delta}R^{\alpha\beta\gamma\delta}$ and
the quadratic Ricci $R_{\alpha\beta}R^{\alpha\beta}$ invariant;
$R_{\alpha\beta\gamma\delta}R^{\alpha\beta\gamma\delta}$
$=C_{\alpha\beta\gamma\delta}C^{\alpha\beta\gamma\delta}$$
+2\,R_{\alpha\beta}R^{\alpha\beta}$$-\frac{1}{3}R^2.$ The
decomposition of the Riemann tensor into its Lorentzian irreducible
parts gives rise to the conformal Weyl (CW) tensor
$C_{\alpha\beta\gamma\delta}$, the traceless Ricci (TR) tensor
$S_{\alpha\beta}$, and the scalar curvature $R$. Petrov
classification gives a full account of the invariant properties of
the CW tensor, see \cite{KramerStephani03}. Pleba\'nski
classification \cite{Plebanski64} deals with the algebraic
properties of $S_{\alpha\beta}$. The first step in the study of the
regularity of solutions is the analysis of the invariant properties
of the TR tensor ${S^\mu}_\nu={R^\mu}_\nu-{R}{\delta^\mu}_\nu /{4}$
due to its relation with the energy--momentum (EM) tensor
${T^\mu}_\nu$, namely, ${S^\mu}_\nu={T^\mu}_\nu-{T}{\delta^\mu}_\nu
/{4}$. Next, one establishes the conditions that allow for
regularity of the CW tensor invariants and finally, the behavior of
the scalar curvature. For the SSS metric there are only three
independent Riemann curvature tensor invariants; the CW tensor
eigenvalue $\Psi_2$, the TR tensor eigenvalue $S$,
 and the scalar $R$.

This article is organized as follows: Section \ref{General} is
devoted to the general description of NLE. In section
\ref{EinsteinEQ}, the Einstein--NLE equations for a SSS metric are
derived explicitly together with the Ricci eigenvalue $S$, the
scalar curvature $R$, and the Weyl $\Psi_2$ invariant. In section
\ref{RegCondN}, see \ref{firstapproach} too, the conditions for the
regularity at the center of an electric NLE SSS metric are
established and a theorem is formulated: Regular NLE SSS electric
solutions, characterized by the Riemanian invariant behavior
$\lim_{\rightarrow 0}\{\Psi_2,S,R\}\rightarrow \{0,0,(0,4\Lambda+
4\mathcal{L}(0))\}$, are regular at the center $r=0$, if and only if
(necessary and sufficient) the metric function $Q(r)$ and the
electric field $q_0F_{rt}=:\mathcal{E}$ behave as $\{Q,\dot Q,\ddot
Q\}\rightarrow\{0,0,2\}$
 and $\{\mathcal{E},\dot\mathcal{E},\ddot\mathcal{E}\}\rightarrow\{0,0,0\}$,
 as $r\rightarrow 0$, i.e., the electric field and its first and
 second order derivatives vanish at the center
 where the metric asymptotically approaches to
the flat or conformally flat de Sitter--Anti de Sitter (dS--AdS)
spacetimes. Section \ref{WECDC} deals with the study of weak and
dominant energy conditions for electrically charged NLE SSS metrics,
the relation Lagrangian--energy density at the center
$\mathcal{L}(0)=\mu(0)$ is established. In section \ref{LinearQ} the
linear superposition of SSS electric solutions of the Einstein--NLE
equations is studied in detail. Section \ref{magnetic} deals with
magnetic metrics characterized by a singular magnetic field
invariant and (possible) regular curvature invariants. In section
\ref{superposition} the general linear superposition property of the
Einstein--pure electric and pure magnetic--solutions is shown. This
article ends with some Final Remarks.

\section{General nonlinear electrodynamics}\label{General}

We follow the standard notation and conventions presented in
\cite{KramerStephani03}. Electrodynamics of any kind are constructed
on the electromagnetic anti--symmetric field tensor  $F_{\mu\nu}$
and its dual tensor ${F^\star}_{\alpha\beta}$
\begin{eqnarray}
F_{\mu\nu}=2A_{[\nu,\mu]},\,
{F^\star}_{\alpha\beta}=\epsilon_{\alpha\beta\mu\nu}F^{\mu\nu},
\end{eqnarray}
where $\epsilon_{\alpha\beta\mu\nu}$ is the totally anti-symmetric
Levi--Civita pseudo--tensor. These tensors determine the invariants
$\mathcal{F}$ and $\check{\mathcal{G}}$, namely
\begin{eqnarray}
 \mathcal{F}=F_{\mu\nu}F^{\mu\nu}/4,\,\,
 \check{\mathcal{G}}=F_{\mu\nu}{F^\star}^{\mu\nu}/4.
\end{eqnarray}
The Einstein gravitational theory coupled to a general nonlinear
electrodynamics is derived from the Riemann--Hilbert action
constructed with the curvature scalar $R$, the Lagrangian
$\mathcal{L}=\mathcal{L}(\mathcal{F},\check{\mathcal{G}})$, and a
$\Lambda$--term, if any,
\begin{eqnarray}
 S= \int {\sqrt{-g}(R-\mathcal{L}-\Lambda)}d^{\,4} x.
\end{eqnarray}
The Einstein equations arising from the variation of this action are
\begin{eqnarray}\label{Einsteina}
{E^\mu}_{\nu}:={R^\mu}_{\nu}-R\,{\delta^\mu}_{\nu}/2
+\Lambda\,{\delta^\mu}_{\nu}-\kappa{T^\mu}_{\nu}=0,
\end{eqnarray}
where the energy--momentum electromagnetic tensor, for
$\mathcal{L}(\mathcal{F},\breve{\mathcal{G}})$, is given by
\begin{eqnarray}\label{EMTen}
{T}_{\mu \nu}=-\mathcal{L}\,{g}_{\mu \nu} + \frac{d\mathcal{L}}{d
\mathcal{F}}\,{F}_{\mu \sigma}
{{F}_{\nu}}^{\sigma}+\frac{d\mathcal{L}}{d
\check{\mathcal{G}}}\,{F}_{\mu \sigma} {{F^\star}_{\nu}}^{\sigma}.
\end{eqnarray}
The electromagnetic field equations are:\\the Bianchi identity
\begin{eqnarray}\label{EMTen}
\,F_{[\mu\,\nu,\,\lambda]}=0\,\equiv {{F^\star}^{\mu\nu}}_{;\nu}=0,
\end{eqnarray}
and the field equations
\begin{eqnarray}\label{EMTen}&&
(\frac{d\mathcal{L}}{d \mathcal{F}}\,{F}^{\mu
\nu}\,+\,\frac{d\mathcal{L}}{d
\check{\mathcal{G}}}\,{{F^\star}}^{\mu\nu})_{;\nu}=0
\nonumber\\&&\equiv\,[\sqrt{-g}(\frac{d\mathcal{L}}{d
\mathcal{F}}\,{F}^{\mu \nu}\,+\frac{d\mathcal{L}}{d
\check{\mathcal{G}}}\,{{F^\star}}^{\mu\nu})]_{,\nu}=0.
\end{eqnarray}
In this work we shall restrict the study to the case
$\mathcal{L}=\mathcal{L}(\mathcal{F})$; the derivative of
$\mathcal{L}$ with respect to $\mathcal{F}$ is denoted by $
\mathcal{L}_\mathcal{F}: =\frac{d\mathcal{L}}{d\mathcal{F}}$. Recall
that Maxwell theory is based on a linear relation between the
Lagrangian $\mathcal{L}$ and the electromagnetic invariant
$\mathcal{F}=F_{\mu\nu}F^{\mu\nu}/4$, such that
$\mathcal{L}=\mathcal{F}\Rightarrow \mathcal{L}_\mathcal{F}=1$ in
the whole spacetime. For physically relevant nonlinear
electrodynamics one may impose on $\mathcal{L}$ and $\mathcal{F}$
the Maxwell weak field limits or ``Maxwell asymptotic'':
\begin{eqnarray}\label{EMTeMax}
 \mathcal{L}\rightarrow \mathcal{F},\,
\mathcal{L}_\mathcal{F}\rightarrow 1, \,\text{for small}
\,\mathcal{F},\,( \text{or as}
 \,\mathcal{F}\rightarrow 0),
\end{eqnarray}
in agreement with Born and Infeld \cite{BornInfeld1934}, see also
\cite{SalazarGarciaPleb1}, i.e., in a weak field limit
$\mathcal{F}$: $\mathcal{F}_{NLE}\rightarrow\mathcal{F}$, and
$\mathcal{L}_{NLE}\rightarrow\mathcal{L}$, such that
$\mathcal{L}\rightarrow\mathcal{F}
;\,\mathcal{L}_{\mathcal{F}}\rightarrow  1 $. Moreover, the NLE
theory ought to fulfill the energy conditions: the local energy has
to be positive and the energy flux has to be carried by a timelike
four vector.

\section{Einstein--NLE equations for a static spherically symmetric
metric}\label{EinsteinEQ}

In this section, for a SSS metric, we derive the Einstein equations
coupled to NLE determined for a Lagrangian $\mathcal{L}$ depending
on the electromagnetic invariant $\mathcal{F}$. Also the curvature
invariants for the SSS metric are determined explicitly. The SSS
metric, restricted only to electrodynamics or vacuum due to the
generalized Birkhoff theorem, can be given in Schwarzschild
coordinates $\{\theta, r, \phi, t\}$ by
\begin{eqnarray}\label{metric}
ds^2 = r^2 d\theta^2+\frac{r^2}{Q(r)}dr^2 +r^2\sin^2\theta
\,d\phi^2-\frac{Q(r)}{r^2}dt^2.
\end{eqnarray}
The Einstein equations coupled to a matter field tensor $T_{\mu\nu}$
and a cosmological constant $\Lambda$, $\kappa=1$, are
\begin{eqnarray}\label{EINSEQ
}E_{\mu\nu}:=G_{\mu\nu}+ \Lambda \,g_{\mu\nu}-T_{\mu\nu}=0.
\end{eqnarray}
The evaluation of the Einstein tensor ${G^\mu}_{\nu}$, and the
curvature scalar $R$, for the metric (\ref{metric}), yields
\begin{eqnarray}
&&{G^\mu}_{\nu}={G^\theta}_{\theta}\left({\delta}^\mu_{\theta}\,{\delta}^\theta_{\nu}
+{\delta}^\mu_{\phi}\,{\delta}^\phi_{\nu}\right)
+{G^t}_{t}\left({\delta}^\mu_{r}\,{\delta}^r_{\nu}
+{\delta}^\mu_{t}\,{\delta}^t_{\nu}\right), \nonumber\\&&
{G^\theta}_\theta=\frac{1}{2}\,\frac{\ddot{Q}}{{r}^{2}}
-\frac{\dot{Q}}{{r}^{3}} +{ \frac {Q }{{r}^{4}}},\, {G^t}_t
=\frac{\dot{Q}}{{r}^{3}}  - \frac {Q }{{r}^{4}} -\frac
{1}{{r}^{2}},\nonumber\\&&
R=\frac{2}{{r}^{2}}-\frac{\ddot{Q}}{{r}^{2}},\,\dot{Q}:=\frac{d}{dr}Q,
\end{eqnarray}
where by means of ``\,\,$\dot{}$\,\,'' is denoted the derivative
with respect to $r$, $\dot{}=\frac{d}{dr}$. It allows, from the
point of view of the eigenvalue problem, for two different double
eigenvalues
$\lambda_\theta=\lambda_\phi={G^\theta}_\theta={G^\phi}_\phi$, and
$\lambda_r=\lambda_t={G^t}_t={G^r}_r$. Thus the related
energy--momentum tensor may describe electrodynamics, see for
instance Lichnerowicz \cite{LichBook1955}. Another way to arrive at
this conclusion is by means of the search for the eigenvalues of the
traceless Ricci tensor
${S^\mu}_\nu={R^\mu}_\nu-\frac{R}{4}{\delta^\mu}_\nu$, which amounts
to
\begin{eqnarray}\label{Sein}
&&{S^\mu}_{\nu}={\it
S}\left({\delta}^\mu_{\theta}\,{\delta}^\theta_{\nu}
-{\delta}^\mu_{r}\,{\delta}^r_{\nu} +{\delta}^\mu
_{\phi}\,{\delta}^\phi_{\nu}
-{\delta}^\mu_{t}\,{\delta}^t_{\nu}\right)
,\nonumber\\
&& {\it S}=\,\frac{\ddot{Q}}{4\,{r}^{2}} -\frac{\dot{Q}}{{r}^{3}} +{
\frac {Q}{{r}^{4}}} +\frac {1}{2\,{r}^{2}},
\end{eqnarray}
$({S^\mu}_\nu)={\it S}\,{\text{diag}} \left( 1,-1,1,-1 \right)$.
Remarkable is the relation ${G^\theta}_{\theta}-{G^t}_{t}=2 {\it
S}$. Therefore, according to the Pleba\'nski \cite{Plebanski64}
classification of matter tensors; this algebraic structure
corresponds to electrodynamics, $[(11)(1,1)]\sim[2\text{ S}-2\text{
T}]_{(11)}$, no matter if it is nonlinear or linear (Maxwell), see
also, Stephani et al. \cite{KramerStephani03}, Chapter 5, and
Chapter 13, \S 13.4. Consequently, static spherically symmetric
metrics of the form (\ref{metric}) allow, besides the vacuum with
$\Lambda$ solutions, $[(111,1)]\sim[4\text{ T}]_{(1)}$, only
electrodynamics solutions to Einstein equations coupled to linear
\cite{LichBook1955} (Maxwell) or nonlinear electrodynamics
\cite{Plebanski64}.

In a certain sense, we are facing a theorem of uniqueness of classes
of electrodynamics solutions; for each electromagnetic invariant
Lagrangian $\mathcal{L}$, related to the electromagnetic invariant
$\mathcal{F}$, the solution is unique. Due to the established
existence of two double eigenvalues different in pairs there is no
room for fluids; any attempt to accommodate in the above
Schwarzschild metric (\ref{metric}) other kind of fields, different
from electrodynamics or vacuum, is spurious, although in the
literature one finds ''solutions''--let us call them better
space--time models--for anisotropic fluids. This is the reason why
we avoid to use the fluid terminology in a place where there is no
room for it.

\subsection{Nonlinear electrodynamics for $\mathcal{L}(
\mathcal{F})$}

This metric structure allows for an energy--momentum (EM) tensor
$T_{\mu\nu}$ associated to the electromagnetic field (EF) tensor
${F_{\mu\nu}}$, $F_{\mu\nu}=2 {A}_{[\mu,\nu]}$. For Einstein--NLE
equations with $\mathcal{L}(\mathcal{F})$, the only non vanishing EF
components are the electric  $ F_{rt}$ and magnetic $F_{\theta\phi}$
fields, therefore the electromagnetic field tensors are
\begin{eqnarray}{F_{\mu\nu}}=2 F_{rt} {\delta^r}_{[\mu}{\delta^t}_{\nu]}+2
F_{\theta\phi} {\delta^\theta}_{[\mu}{\delta^\phi}_{\nu]}
\end{eqnarray}
\begin{eqnarray}{F^{\mu\nu}}=-2 F_{rt} {\delta^\mu}_{[r}{\delta^\nu}_{t]}+2
{\frac {F_{\theta\phi}
 }{{r}^{4}  \sin^{2} \left( \theta \right)
  }} {\delta^\mu}_{[\theta}{\delta^\nu}_{\phi]}.
\end{eqnarray}
Consequently, the electromagnetic invariant
$\mathcal{F}={F_{\mu\nu}}{F^{\mu\nu}}/4$ is given by
\begin{eqnarray}\label{invariant}&&
\mathcal{F}=-\frac{1}{2}\, \left( F_{rt}
 \right) ^{2}+\frac{1}{2}\,
 {\frac { \left( { F_{\theta\phi}} \right) ^{2}}{{r}^{4}  \sin^2 \left( \theta \right)
 }}={\mathcal{F}}_e+{\mathcal{F}}_m,\nonumber\\&&
 {\mathcal{F}}_e=-\frac{1}{2}\, \left( F_{rt}
 \right) ^{2},\,{\mathcal{F}}_m
 =\frac{1}{2}\,
 {\frac { \left( { F_{\theta\phi}} \right) ^{2}}{{r}^{4}  \sin^2 \left( \theta \right)
 }}.
\end{eqnarray}
The Bianchi identities $F_{[\mu\nu;\alpha]}=0$ and the EF equations,
\begin{eqnarray*}(\frac{d\mathcal{L}}{d\mathcal{F}}{F^{\mu\nu}})_{;\nu}=0
\Rightarrow(\sqrt{-g}\mathcal{L}_{\mathcal{F}}{F^{\mu\nu}})_{,\nu}=0\,
\end{eqnarray*}
\begin{eqnarray} \simeq\frac{\partial}{{\partial x^\nu}}\left(
{r}^2\sin{\theta}\mathcal{L}_{\mathcal{F}} \,{F}^{\mu\nu}\right)=0,
\end{eqnarray}
where\,$\frac{d \mathcal{L}}{d \mathcal{F}}=:
\mathcal{L}_{\mathcal{F}}$, lead to the electric field equation
\begin{eqnarray}\label{electric}
\mathcal{L}_{\mathcal{F}}\,{F}_{{r}t}=-\frac{q_0}{{r}^2},\,\,
\mathcal{F}_e=-\frac{1}{2}\left(F_{rt}\right) ^{2},
\end{eqnarray}
where $q_0$ is a constant related with the electric charge, and to
the magnetic field equation
\begin{eqnarray}\label{magnetic}
 F_{\theta\phi}=h_0\sin{\theta},\,{\mathcal{F}}_m=\frac{1}{2}\,\frac {
 h_0^{2}}{{r}^{4}},
\end{eqnarray}
where $h_0$ is a magnetic constant. Moreover, the electric field
equation (\ref{electric}) can be given alternatively as
\begin{eqnarray}\label{electric2}
&& {{F}_{{r}t}({r}^2\,\frac{d \mathcal{L}}{dr}
-q_0\frac{d}{d{r}}{F}_{{r}t})}- 2\frac{q_0h_0^2}{r^5}=0.
\end{eqnarray}
The EM tensor ${T^{\mu}}_{\nu}$ and its trace $T$ can be given as
 \begin{eqnarray} {T^{\mu}}_{\nu}&&= \left(-\mathcal{L}
+\mathcal{L}_{\mathcal{F}}{h_0^2}/{{r}^4}\right) \left(
\delta^{\mu}_{\theta}\delta^{\theta}_\nu
+\delta^{\mu}_{\phi}\delta^{\phi}_\nu
 \right)\nonumber\\&& - \left(\mathcal{L}
+\mathcal{L}_{\mathcal{F}} { F_{{r}t}}^2\right)
 \left(\delta^{\mu}_{r}\delta^{r}_\nu
+\delta^{\mu}_{t}\delta^{t}_\nu
 \right),\nonumber\\&&
T := {T_{\mu}}^{\mu}
=-4\mathcal{L}+4\mathcal{L}_{\mathcal{F}}\left(\mathcal{F}_e
+\mathcal{F}_m\right),
 \end{eqnarray}
from where it is apparent that its eigenvalues fulfill
$\lambda_\theta=\lambda_\phi$, and $\lambda_r=\lambda_t$, see\,
\cite{LichBook1955}.

Notice that in the general electromagnetic (dyonic) case the
derivative $\frac{d\mathcal{L}}{d\mathcal{F}}=\frac{d
\mathcal{L}}{d{r}}/\frac{d \mathcal{F}}{d{r}}$, $-\frac{d
\mathcal{F}}{d{r}}={F_{{r}t}\frac{dF_{{r}t}}{d{r}}+2\frac{h_0^2}{{r}^5}}
$, which makes the Einstein equations quite involved.

The nontrivial components of the Einstein--NLE equations are
$E^r_r=E^t_t$ and $E^\theta_\theta=E^\phi_\phi$, hence the Einstein
equations reduce to
\begin{eqnarray}\label{EQET1}
&&E^t_t =\frac{\dot{Q}}{{r}^{3}} -{ \frac {Q }{{r}^{4}}} -\frac
{1}{{r}^{2}}+\Lambda+\mathcal{L}+\frac{d \mathcal{L} }{d
\mathcal{F}} \left(F_{rt}\right) ^{2}=0,\nonumber\\&&
E^\phi_\phi
=\frac{\ddot{Q}}{2\,{r}^{2}} -\frac{\dot{Q}}{{r}^{3}}
 +{ \frac {Q }{{r}^{4}}} +\Lambda+\mathcal{L}+\frac{d \mathcal{L} }{d
\mathcal{F}}\,\frac{h_0^2}{r^4} =0,
\end{eqnarray}
Using the EF equation
 $\mathcal{L}_{\mathcal{F}}\,{F}_{{r}t}=-q_0/{r}^2$, one
may isolate, via subtraction, ${q_0} F_{rt} $ and $\mathcal{L}$ as
\begin{eqnarray}\label{Einstein1Fyt2x}&&
{q_0}\, F_{{r}t} \left( {r} \right) =-\frac{\ddot{Q}}{2} +
2\,\frac{\dot{Q}}{{{r}}} -2\,\frac {Q}{{r}^2}-1+\frac { h_0^{2}}{
{{r}}^2 }{\mathcal{L}}_{\mathcal{F}},\nonumber\\&& \mathcal{L}({r})
=-\frac{\ddot{Q}}{2{{r}}^{2}} +\frac {\dot{Q}}{{{r}}^ {3}}
-\frac{Q}{{{r}}^{4}} -\Lambda +{\frac { h_0^{2} }{{{r}}^{4}
}}{\mathcal{L}}_{\mathcal{F}}.
 \end{eqnarray}
to determine $F_{{r}t}({r})$, $\mathcal{L}(r)$, and the structural
function $Q({r})$.

The EM conservation equation ${T^{\mu\nu}}_{;\nu}=0$ leads to the
condition
\begin{eqnarray}\label{ConserT2E}&&{r}^2\frac{d}{d{r}}
\mathcal{L}({r})-\frac{d}{d{r}} (q_0{F_{{r}t}} \left( {r}
\right))+2\,\frac {h_0^{2} }{{{r}}^{3}} {\mathcal{L}}_\mathcal{F}=0,
\end{eqnarray}
which becomes an identity--Bianchi identity--by using the Einstein
equations (\ref{Einstein1Fyt2x}). The substitution of $q_0F_{rt}$
and $\mathcal{L}$ from (\ref{Einstein1Fyt2x}) into (\ref{electric2})
gives the differential equation of (\ref{electric}) multiplied by
$2h_0^2/r^3$, therefore, the only differential equations to be
integrated are the (\ref{Einstein1Fyt2x}) ones.

\subsection{ Euler equations
and their solutions in electric NLE for given $\mathcal{E}= q_0
F_{rt}$}\label{Eulerqsec1}
 The Einstein--NLE field equations
(\ref{Einstein1Fyt2x}) can be analyzed from the point of view of
Euler equations for the function $Q(r)$. These equations, in the
electric case, can be written as
\begin{eqnarray}\label{Einstein1Fyt2xcF}&&
D_fQ:=\left(r^2\frac{d^2}{dr^2}-4r\frac{d}{dr}+4\right)Q=
-2r^2-2r^2{ q_0}\, F_{rt} ,\nonumber\\&&
D_LQ:=\left(r^2\frac{d^2}{dr^2}-2r\frac{d}{dr}+2\right)Q=
-2r^4\Lambda-2r^4\mathcal{L} .
 \end{eqnarray}
Each one of these equations is a non homogeneous Euler equation. The
solutions of a homogeneous Euler equation are searched as power of
$r$, i.e., $r^k$. The solution for the non--homogeneous Euler
equation is determined by using the method of variation of
parameters \footnote {A linear second order equation
$$p(r)\ddot{y}+q(r)\dot{y}+s(r)y(r)=f(r)$$
allows two independent homogeneous solutions $y_1(r)$ and $y_2(r)$,
which determine the homogeneous solution $y_{h}=A y_1(r)+B y_2(r),$
the non homogeneous solution is sought in the form $y_{nh}=A(r)
y_1(r)+B(r) y_2(r)$, under the conditions $\dot{A}(r)
y_1(r)+\dot{B}(r) y_2(r)=0$, its substitution into the non
homogeneous equation yields $\dot{A}(r) \dot{y_1}(r)+\dot{B}(r)
\dot{y}_2(r)=f(r)/p(r)$, thus
$\dot{A}=-\frac{f}{p}\frac{y_2}{W(y_1,y_2)} $ and
$\dot{B}=\frac{f}{p}\frac{y_1}{W(y_1,y_2)} $, where
${W(y_1,y_2)}=y_1\dot{y}_{2}-\dot{y}_{1}y_2$, integrating these
equations one gets the non--homogeneous solution as
$$y_{nh}=-y_1\int{\frac{f}{p}\frac{y_2}{W(y_1,y_2)} dr}
+y_2\int{\frac{f}{p}\frac{y_1}{W(y_1,y_2)} dr},$$ which, added to
the homogeneous solution $y_{h}$, with constants $A$ and $B$, gives
the general solution.}. For instance, considering the field $F_{rt}$
as a given function, one integrates the first Euler equation for
$Q$, with homogeneous solutions proportional to $r^4$ and $r$,
namely $C_1r+C_4r^4$, via the variations of parameters, arriving at
the general linear integral solution
\begin{eqnarray}\label{EulerQ}&&
{Q(r)}=Q_K+Q(\mathcal{E}),\,\, \mathcal{E}(r):= q_0\,F_{rt} \left( r
\right), \nonumber\\&& \,Q(\mathcal{E}):=-\frac{2}{3}\,{r}^{4}\int
\! {\frac {\mathcal{E}\left( r \right) }{{r}^{3}}}{dr}
+\frac{2}{3}\,r\int \!\mathcal{E}\left( r \right) {dr},
\,\,\nonumber\\&& Q_K:={r}^{2}-2{\it
m}\,r-\frac{\Lambda_e}{3}\,{r}^{4},\Lambda_e=\Lambda+\mathcal{L}(0).
\end{eqnarray}
The constants have been chosen as $C_1=-2\,m$ and $C_4=-\Lambda_e
/3$, where $\Lambda_e=\Lambda+\mathcal{L}(0)$, hence $\Lambda_e$
stands for an effective $\Lambda_e$--term; for correspondence with
the vacuum plus cosmological constant $\Lambda$--Kottler solution
\cite{Kottler}, $\Lambda_e=\Lambda$. In this parametrization, the
function $Q_K$ corresponds to the Kottler--like solution to vacuum
plus cosmological constant Einstein equations, i.e., the
``Schwarzschild--de Sitter--Kottler'' solution, where the parameter
$m$ can be thought of as the ``Schwarzschild mass''.

The corresponding Lagrangian function $\mathcal{L}$ , arising from
the substitution of $ Q(r)$ from (\ref{EulerQ}) into
(\ref{Einstein1Fyt2xcF}) amounts to
\begin{eqnarray}\label{EulerQL}
\,\mathcal{L}(\mathcal{E})= \mathcal{L}(0)+2\,\int_0^r \!{\frac
{\mathcal{E}(r) }{{ r}^{3}}{dr}}+\frac {\mathcal{E} ( r) }{{r}^{2}},
\end{eqnarray}
where it has been replaced $ \Lambda_e-\Lambda$ by $\mathcal{L}(0)$.
The integral is assumed to be definite; for an indefinite integral
one has
\begin{eqnarray}\label{EulerQLdef} \,\mathcal{L}(\mathcal{E}(r))=
2\,\int^r \!{\frac {\mathcal{E}(x) }{{ x}^{3}}{dx}}+\frac
{\mathcal{E} ( r) }{{r}^{2}}.
\end{eqnarray}
For regular solutions, $\lim_{\rightarrow 0}S(=-\frac {\mathcal{E} (
r) }{{r}^{2}})\rightarrow 0 $, the integration constant
$\mathcal{L}(0)=\int^0 \!{\frac {\mathcal{E}(r) }{{
r}^{3}}{dr}}$--meaning integral evaluated at zero.

We shall see in the subsection \ref {firstapproach}, and section
\ref{WECDC}, related with the regularity conditions, that the
electrodynamics trace--curvature scalar relation (\ref{CruR}) at the
center gives rise to the $\mathcal{L}(0)$ constant.

Finally, still talking about the integration process for electric
NLE solutions, there exist another alternative approach related with
the electric field equation, or, equivalently, with the
energy--momentum conservation equation (\ref{ConserT2E}). In the
electric case, the substitution of a given function
$\mathcal{L}=\mathcal{L}(\mathcal{F})$ into the electric field
equation,
\begin{eqnarray*}\label{ConserT2Ef}&&
\mathcal{L}_\mathcal{F}{ F_{{r}t}}=-{q_0}/{r^2},
\end{eqnarray*}
taking into account that $\mathcal{F}=-{ F_{{r}t}}^2/2$, leads a
relation between $F_{rt}$ and $r$, in general a transcendent one;
thus, being lucky, one gets explicitly $q_0F_{rt}=\mathcal{E}(r)$ as
a function of $r$. Next, the substitution of this $\mathcal{E}(r)$
into (\ref{EulerQ}) gives the metric function $Q(r)$. }

\section{The three curvature invariants of the SSS metric: $S$,
$\Psi_2$, and R}

It is known that the necessary and sufficient regularity conditions
of an arbitrary spacetime are determined by the regular behavior {of
its curvature invariants everywhere in the whole spacetime. In this
article we focus on the second order curvature invariants}. Lake and
Musgrave \cite{LakeMusgrave94} studied the regularity of static
spherically symmetric, cylindrically symmetric and plane symmetric
spacetimes at the origin; although a set of specific invariants was
evaluated for each class, the authors acknowledged in \S 2 that
``...the invariants associated with the metric (1) can be evaluated
relatively quickly, whereas they are too cumbersome to reproduce
here in their entirety.''

Later, in 1997, Zakhary and McIntosh \cite{McIntoshZakhary97}
determined a complete set of independent curvature invariants for
spacetimes of Lorentzian signature; in the abstract of their 43
pages publication, one reads: ``There are at most 14 independent
real algebraic invariants of the Riemann tensor in a four
dimensional Lorentzian space. In the general case, these invariants
can be written in terms of four different types of quantities: $R$,
the real curvature scalar, two complex invariants $I$ and $J$ formed
from the Weyl spinor, three real invariants $I_6$, $I_7$ and $I_8$
formed from the trace-free Ricci spinor and three complex  mixed
invariants $K$, $L$ and $M$.'' Moreover, in section 10, these
authors reviewed previously reported (incomplete as a consequence of
ZM \cite{McIntoshZakhary97}) sets of Riemann invariants. The ZM set
of invariants, \cite{McIntoshZakhary97}(84), beside the scalar
curvature $R$, in terms of the Weyl spinor $\Psi_{ABCD}$ and
traceless Ricci spinor $\Phi_{AB\dot{C}\dot{D}}$, are defined as
follows:
\begin{eqnarray}&&
I=\frac{1}{6}\Psi_{ABCD}\Psi^{ABCD},\nonumber\\&&
J=\frac{1}{6}\Psi_{ABCD}{\Psi^{CD}}_{EF}\Psi^{EFAB},\nonumber\\&&
I_6=\frac{1}{3}\Phi_{AB\dot{C}\dot{D}}\Phi^{AB\dot{C}\dot{D}},\nonumber\\&&
I_7=\frac{1}{3}\Phi_{AB\dot{C}\dot{D}}
{{{{\Phi^{B}}_{E}}^{\dot{D}}}}_{{\dot{F}}}\Phi^{AE\dot{F}\dot{C}},\nonumber\\&&
I_8=\frac{1}{6}\Phi_{AB\dot{C}\dot{D}}{{{{\Phi^{B}}_{E}}^{\dot{C}}}}_{{\dot{F}}}
\,\Phi^{AG\dot{H}\dot{D}}{{{{\Phi^{E}}_{G}}^{\dot{F}}}}_{{\dot{H}}}\nonumber\\
&&{}{}+
\frac{1}{6}\Phi_{AB\dot{C}\dot{D}}{{{{\Phi^{B}}_{E}}^{\dot{C}}}}_{{\dot{F}}}
\Phi^{AG\dot{H}\dot{F}}{{{{\Phi^{E}}_{G}}^{\dot{D}}}}_{{\dot{H}}}
,\nonumber\\&&
K=\Psi_{ABCD}{\Phi^{CD}}_{\dot{E}\dot{F}}\Phi^{AB\dot{E}\dot{F}}.
\end{eqnarray}
The spinorial formulation of the general relativity theory was done
by Penrose and Rindler \cite{Penrose86}, for a brief description of
spinors see \cite{KramerStephani03}, Chapter 3.

Recently, Torres and Fayos \cite{TorresFayos17} studied the
re\-gula\-rity of Kerr-like black holes of Petrov type D. Based on a
theorem by Zakhary and  McIntosh \cite{McIntoshZakhary97}, they
formulated their Proposition~3: `` The algebraically second order
set of invariants for a Petrov type D spacetime and Segre type
$[(1,1)(11)]$ is $\{R,I,I_6,K\}$,'' where, in tensorial description,
these quantities are defined as:
\begin{eqnarray}\label{Invariants}&&
R={R^\alpha}_\alpha,\nonumber\\&&
I_6=\frac{1}{12}{S_\alpha}^\beta{S_\beta}^\alpha,\nonumber\\&&
I=\frac{1}{24}{\tilde{C}}_{\alpha\beta\gamma\delta}{\tilde{C}}^{\alpha\beta\gamma\delta},
\,{\star{C}}_{\alpha\beta\gamma\delta}:=\epsilon_{\alpha\beta\mu\nu}{{C}^{\mu\nu}}_{\gamma\delta},\nonumber\\&&
\,{\tilde{C}}_{\alpha\beta\gamma\delta}:=\frac{1}{2}({{C}}_{\alpha\beta\gamma\delta}+
i {\star{C}}_{\alpha\beta\gamma\delta}), \nonumber\\&&
K=\frac{1}{4}{\tilde{C}}_{\alpha\beta\gamma\delta}{S}^{\alpha\delta}{S}^{\beta\gamma}.
\end{eqnarray}
For a Petrov type $D$, $I^3=J^2$, and a non isotropic
electromagnetic field, these invariants become:
\begin{eqnarray}\label{Invariants}&&
R={R^\alpha}_\alpha,\nonumber\\&& I_6=C_{I_6}\,{S^2},\nonumber\\&&
I=C_I\,{\Psi_2}^2,\nonumber\\&&
 K= C_K\,{\Psi_2}\,{S}^2.
\end{eqnarray}
where $C_T, T=\{I_6,I,K\}$ stand for numerical coefficients, whose
specific values depend on the adopted conventions.

Summarizing, one arrives at the proposition: The necessary and
sufficient conditions for the regularity at the center of a Petrov
type $D$ spacetime coupled to a non--null electromagnetic field of
Segre type $[(11)(1,1)]$ are determined by the finiteness of the
independent eigenvalue $\Psi_2$ of the Weyl tensor matrix, the
eigenvalue $S=\Phi_{11}/2$ of the traceless Ricci tensor matrix, and
the curvature scalar $R$. In particular this proposition holds for
the studied here static spherically symmetric metric, which is
always of Petrov type D and allows for NLE with electromagnetic
field of Segre type $[(11)(1,1)]$.

We make use of the null tetrad formalism to determine the curvature
Weyl and traceless Ricci tensor quantities for the SSS metric
described as
\begin{eqnarray}
g=2{\bf{e^1}\bf{e^2}}-2{\bf{e^3}\bf{e^4}}=g_{ab}{\bf{e^a}\bf{e^b}},\,\,\,
\bf{e^a}&&={e^a}_\mu \,\bf{dx}^{\mu},
\end{eqnarray}
\begin{displaymath}
 \left.\begin{array}{cc}{
 \bf {e^{1}}}\\  {\bf{e^{2}}}
\end{array}\right\}
= \frac{1}{\sqrt{2}}\,\left({r\,{\bf d \theta}}\pm
i\,r\,\sin{\theta} {\bf d \phi} \right),
\end{displaymath}
\begin{displaymath}
 \left.\begin{array}{cc}{
 \bf {e^{3}}}\\  {\bf{e^{4}}}
\end{array}\right\}
= \frac{1}{\sqrt{2}}\,\left(\frac{r\,{\bf
dr}}{\sqrt{Q(r)}}\pm\,\frac{{\sqrt{Q(r)}}{\bf dt}}{r}\right),
\end{displaymath}

The tetrad transformation matrix is given by $({h^{\bf
a}}_\mu)=([{\bf e^a}_\mu])$, ${\bf a}=1,2,3,4,$ and
$\mu=\{\theta,r,\phi,t\},$ in a row arrangement.\\
The invariant characterization of the algebraic properties of the
gravitational--matter field, as it has been widely detailed above,
begins with the determination of the eigenvalues of the TR tensor
${S^\mu}_\nu$. In the studied case ${S^\mu}_\nu$ amounts to
$({S^\mu}_{\nu})=S\,\text{ diag}(1,-1,1,-1)=\text{
diag}(\lambda_\theta,\lambda_r,\lambda_\phi,\lambda_t)$, where
\begin{eqnarray}\label{EigenSratPsi2x}
&&{ S}=\,\frac{\ddot{Q}}{4\,{{r}}^{2}} -\frac{\dot{Q}}{{{r}}^{3}} +{
\frac {Q}{{{r}}^{4}}} +\frac
{1}{2\,{{r}}^{2}}=2\Phi_{11}=S_{12}=S_{34}\nonumber\\&&
=\frac{1}{2}{\mathcal{L}}_{\mathcal{F}}\left( F_{{r}t}^2
+\frac{{h_0}^2}{{r}^4}\right)
=-\frac{q_0}{2\,{r}^2}F_{{r}t}+\frac{1}{2}
{\mathcal{L}}_{\mathcal{F}}\,\frac{h_0^2}{{r}^4}.
\end{eqnarray}
For the SSS metric, the Weyl curvature invariant $\mathcal{C}^2:=
C_{\alpha\beta\gamma\delta}C^{\alpha\beta\gamma\delta}$ amounts to
 $\mathcal{C}^2=48\,{\Psi_2}^2$, where the Weyl tetrad coefficient
 $\Psi_2$ is given by
\begin{eqnarray}\label{curatPsi2}
&&-12\,{r}^4\,\Psi_2={{r}}^{2}\ddot{Q} -6\,{r}\, \dot{Q} +12 \,Q
-2\,{{r}}^{2}.
\end{eqnarray}
From the point of view of Petrov classification, the $\Psi's$ are
related to the eigenvalues $\lambda$ of the eigenbivector equation
$C_{abcd} X^{cd}=\lambda X_{ab},\, \lambda_1+ \lambda_2+
\lambda_3=0.$ According to the Table 4.2 of \cite{KramerStephani03},
the studied metric is of Petrov type D, with eigenvalues
$\lambda_1=\lambda_2=-2\,\Psi_2$, while the remaining $\Psi's$
vanish.\\
The scalar Riemann curvature $R$ is given by
\begin{eqnarray}\label{curvaR}
{{{r}}^{2}}R={2}-{\ddot{Q}},\,\,R=4\mathcal{L}-4\mathcal{L}_{\mathcal{F}}\left(\mathcal{F}_e
+\mathcal{F}_m\right)+4\Lambda.
\end{eqnarray}
The independent curvature invariants for this SSS metric can be
given explicitly as
\begin{eqnarray}\label{curvaRKret1x}&&
I_6=\frac{1}{12}S_{\alpha\beta}S^{\alpha\beta}=3\,S^2,\,2\Phi_{11}=S,\nonumber\\&&
\mathcal{I}=\frac{1}{24}C_{\alpha\beta\gamma
\delta}C^{\alpha\beta\gamma \delta}=2\,\Psi_2^2,\nonumber\\&&
\mathcal{K}=\frac{1}{4}\,C_{\alpha\beta\gamma
\delta}S^{\alpha\delta}S^{\beta\gamma}=4 \Psi_2\,S^2.
\end{eqnarray}
Incidentally, the invariants $\Psi_2$ and $S$ can be considered as
``square roots'' of the quadratic curvature invariants $I_6$ and
$\mathcal{I}$. The Kretschmann quadratic Riemannian invariant
\begin{eqnarray}\label{curvaRKret1}
R_{\alpha\beta\gamma \delta}R^{\alpha\beta\gamma
\delta}=C_{\alpha\beta\gamma \delta}C^{\alpha\beta\gamma
\delta}+2\,S_{\alpha\beta}S^{\alpha\beta}+\frac{1}{6}R^2
\end{eqnarray}
for the SSS metric becomes
\begin{eqnarray}\label{curvaRKret1}&&
R_{\alpha\beta\gamma \delta}R^{\alpha\beta\gamma
\delta}=48\,\Psi_2^2+8\,S^2+\frac{1}{6}R^2.
\end{eqnarray}
Therefore, we succeeded to shift the Riemann invariants'
characterization for static spherically
 symmetric spacetime coupled to NLE fields (of Segre type $[(1,1)(11)])$
 to a linear characterization in terms of a simple set of three
 invariant functions, namely the eigenvalue of the Weyl conformal tensor matrix,
 the eigenvalue of the traceless
 Ricci tensor matrix, and the curvature scalar,
 $\{\Psi_2,\,S,\,R\}$. These three invariant functions depend linearly on the single
 metric function $Q(r)=-r^2g_{tt}$ and its first and second order
 derivatives. Because of their general character, they constitute
 an efficient tool for the determination of general properties of classes of
 spacetimes. Notice that the quadratic Riemann invariant,
 $R_{\alpha\beta\gamma \delta}R^{\alpha\beta\gamma
\delta}$, one of the four Riemannian invariants, is just in the
studied case the sum of the squares of $\Psi_2$, $S$, and $\,R$,
modulo positive numerical factors. Consequently, it is more easy to
evaluate and extract information from $\{\Psi_2,\,S,\,R\}$ than from
the single Kretschmann quadratic Riemannian invariant, which, in any
case, will provide partial information.

\section{Regularity of $\{\Psi_2,\,S,\,R\}$; first
approach}\label{firstapproach}

In this section we shall establish the necessary and sufficient
conditions for the regularity of the curvature invariants
$\{\Psi_2,\,S,\,R\}$ given as
\begin{eqnarray}\label{EigenSratPsi2xY}
&&4r^4{ S}={{r}}^{2}\,{\ddot{Q}} -4r\,{\dot{Q}} +4{Q} +\,2{{r}}^{2},
\end{eqnarray}
\begin{eqnarray}\label{curatPsi2Yo}
&&-12\,{r}^4\,\Psi_2={{r}}^{2}\ddot{Q} -6\,{r}\, \dot{Q} +12 \,Q
-2\,{{r}}^{2},
\end{eqnarray}
\begin{eqnarray}\label{curvaRY1}
{{{r}}^{2}}R={2}-{\ddot{Q}}.
\end{eqnarray}
First let us analyze the behavior of the metric function $Q(r)$.
Assuming the finiteness of curvature scalar $R$ at the center $r=0$,
$R(0)=FQ$, where $FQ$ stands for finite quantity, the equation
(\ref{curvaRY1}), as $r$ approaches to zero, becomes
\begin{eqnarray}&&\label{curvaRYN}
0={2}-\lim_{r\rightarrow 0}{\ddot{Q}}\,\Rightarrow
\,{\ddot{Q}}(0)=2.
\end{eqnarray}
Next, the finiteness of the traceless Ricci tensor eigenvalue $S$,
(\ref{EigenSratPsi2xY}), and the Weyl curvature $\Psi_2$,
(\ref{curatPsi2Yo}), at zero, yields to
\begin{eqnarray}\label{EigenSratPsi2xY2}
\lim_{r\rightarrow 0}Q=0 \Rightarrow Q(0) =0.
\end{eqnarray}
A function $Q(r)$, fulfilling these limiting conditions, which is
also a solution of the $R$--equation for vanishing $R=0$,
$0=2-{\ddot{Q}}(r)$, is given by
\begin{eqnarray}\label{SOLvacS}
Q(r)=r^2-2m\,r,
\end{eqnarray}
which one identifies with the structural function for the vacuum
Schwarzschild solution. Evaluating $\Psi_2$ from
(\ref{curatPsi2Yo}), and $S$ from (\ref{EigenSratPsi2xY}) for Q(r)
from (\ref{SOLvacS}), one gets
$$\Psi_2(r)=\frac{m}{r^3},\,\,S(r)=0.$$
Therefore, for regularity in the limit as $r$ approaches to $0$, the
mass $m$ has to be equated to zero. This function $Q(r)$,
(\ref{SOLvacS}) with $m=0$, as $r\rightarrow 0$, behaves as the flat
metric function
\begin{eqnarray}\label{SOLvacx}
Q(r)=r^2,\,\,\lim_{r\rightarrow 0}\{ Q, \dot Q,\ddot Q\}=\{0,0,2\}.
\end{eqnarray}
Moreover, one may analyze the equation (\ref{curvaRY1}) for the
function $Q(r)$, for small $r$, from point of view of the series
expansion of the curvature $R$, i.e., for its zero order leading
term $R(0)$, namely
\begin{eqnarray}&&\label{curvaRYx0c}
r^2\,R(0)\simeq{2}-{\ddot{Q}}.
\end{eqnarray}
Integrating (\ref{curvaRYx0c}), taking into account that $Q(0)=0$,
one gets
\begin{eqnarray}&&\label{curvaRYxvv}
Q(r)\simeq r^2-2\,m\,r-\frac{R(0)}{12} r^4,\,\,\text{as}
\,r\rightarrow \,0 ,
\end{eqnarray}
which is a Kottler--like metric function; the proper Kottler
function $Q_K=r^2-2\,m\,r-\Lambda r^4/3$ arises from
(\ref{curvaRYxvv}) for $R(0)=4\Lambda$.

Evaluating $\Psi_2$ from (\ref{curatPsi2Yo}), and $S$ from
(\ref{EigenSratPsi2xY}) for Q(r) from (\ref{curvaRYxvv}), one gets
$$\Psi_2\simeq\frac{m}{r^3},\,\,S\simeq0.$$
For a finite behavior of $\Psi_2$, and  of the structural function
$g_{tt}=-Q(r)/r^2$, the ``Schwarzschild mass'' $m$ has to vanish.\\

Consequently, from the geometrical point of view based on the metric
structure, the regularity of the curvature invariants at the center,
$\lim_{{r}\rightarrow 0}\{S,\Psi_2,R\}=\{S(0),0,(0,R(0))\}$,
requires necessarily that the metric function $Q$ has to fulfill the
conditions
\begin{eqnarray}&&\label{RegCOND}
\lim_{{r}\rightarrow 0}\{Q,\dot{Q},\ddot{Q}\}=\{0,0,2\}.
\end{eqnarray}

On the other hand, from the perspective of the electrodynamics, the
remaining equations are related to the field quantities, namely, the
electric field equation
\begin{eqnarray}\label{EigenSratYxab1}
{r}^2F_{{r}t}{\mathcal{L}}_{\mathcal{F}}=-{q_0},
\end{eqnarray}
and the traceless Ricci tensor invariant $S$, related to the field
$\mathcal{F}$ through $S({\mathcal{F}})$,
\begin{eqnarray}\label{EigenSratYxab2}
&&{S} =-{\mathcal{F}}{\mathcal{L}}_{\mathcal{F}},
\end{eqnarray}
which leads, via the field equation (\ref{EigenSratYxab1}), to
\begin{eqnarray}\label{EigenSratYxab}
&&2r^2{\it S}=-q_0F_{{r}t} =:-\mathcal{E}.
\end{eqnarray}
From this last relation it is clear that $\mathcal{E}$ and $S$, as
series expansions, ought to obey
\begin{eqnarray}\label{EigenSrSE}
\mathcal{E}(r)\simeq \mathcal{E}_2 r^2+ \mathcal{E}_3
r^3+...,\,\,S(r)\simeq-\frac{\mathcal{E}_2}{2}-\frac{\mathcal{E}_3}{2}r+...,\end{eqnarray}
for the regularity of ${\it S}$ as $r\rightarrow 0$, otherwise, if
$\mathcal{E}_0\neq0\neq\mathcal{E}_1$, then $$ S\simeq
-\frac{\mathcal{E}_0}{2r^2}-\frac{\mathcal{E}_1}{2r}-\frac{\mathcal{E}_2}{2}+...,$$
hence $\lim_{{r}\rightarrow
0}{\it S}\rightarrow \infty$ as $r\rightarrow 0$.\\
Therefore, for finite $S$ at $0$, one has to have ${\it
S}(0)=-\mathcal{E}_2/2$, and $\lim_{{r}\rightarrow 0}\mathcal{E}=0$,
$$ \lim_{{r}\rightarrow
0}\{\mathcal{E},\dot\mathcal{E},\ddot\mathcal{E}\}
=\{0,0,2\mathcal{E}_2\}=\{0,0,-4S(0)\},$$ as
$r$ approaches to zero.\\
On the other hand, there is still the invariant electric
trace--scalar curvature relation
\begin{eqnarray}\label{CruR}
R(r)=4\Lambda+4\mathcal{L}+4{\it S}.
\end{eqnarray}
For regular solutions, the invariant curvature $R$ ought to be
finite at the origin: $\lim_{{r}\rightarrow 0}\,R= R(0)$, i.e., a
finite quantity $(FQ)$, hence from (\ref{CruR}) one gets
\begin{eqnarray}\label{CruRa}
\lim_{{r}\rightarrow 0}R(r)=R(0)=4\Lambda+4\lim_{{r}\rightarrow
0}\mathcal{L}+4\lim_{{r}\rightarrow 0}{\it S}(r).
\end{eqnarray}
For finite ${\mathcal{L}}$ one has two possibilities: a finite
quantity
\begin{eqnarray}\label{Lag1}
\lim_{{r}\rightarrow 0}\,{\mathcal{L}}
=\mathcal{L}(0),
\end{eqnarray}
which contributes to $Q(r)$ with an
additional $\Lambda$--term, or
$$ \lim_{{r}\rightarrow 0}\,{\mathcal{L}}=0. $$
The class of regular electric fields at the center, with a zero
scalar curvature $R(0)=0$ or an effective cosmological constant
\begin{eqnarray}\label{CruRCOND0}
R(0)= 4\Lambda+4\mathcal{L}(0),
\end{eqnarray}
is determined by the vanishing of
\begin{eqnarray}\label{CruRCOND}
\lim_{{r}\rightarrow 0}\,S(r)=0=S(0)\Rightarrow \mathcal{E}_2 =0,
\end{eqnarray}
where the last relation arises from the series expansions
(\ref{EigenSrSE}). Consequently, the regularity of the curvature
quantities at the center $\lim_{{r}\rightarrow
0}\{\Psi_2,S,R\}=\{0,0,(0,4\Lambda+4\mathcal{L}(0))\}$ takes place
if and only if
\begin{eqnarray}\label{CruRxx}
\lim_{r\rightarrow 0}\{Q,\dot Q,\ddot Q\}=\{0,0,2\},
\end{eqnarray}
together with
\begin{eqnarray}\label{CruRxxy}
\lim_{r\rightarrow
0}\{\mathcal{E},\dot\mathcal{E},\ddot\mathcal{E}\}=\{0,0,0\}.
\end{eqnarray}
It is important to notice the crucial role played by the relation
$S(\mathcal{E})$, (\ref{EigenSratYxab}), and the series expansions
(\ref{EigenSrSE}) in the determination of the regularity conditions,
which sent the relation $S(\mathcal{F})$, (\ref{EigenSratYxab2}), to
occupy an irrelevant secondary place; the relation $S(\mathcal{F})$,
(\ref{EigenSratYxab2}), does not contribute directly to the
establishing of the regularity conditions. Some authors used both
equations simultaneously to derive the searched regularity
conditions, this procedure may lead
to inconsistencies.\\

Isolating the gradient ${\mathcal{L}}_{\mathcal{F}}$ from the
electric field equation (\ref{EigenSratYxab1}) one gets
\begin{eqnarray}\label{EigenSratYxab1a}
{\mathcal{L}}_{\mathcal{F}}=-\frac{q_0}{{r}^2F_{rt}} ,
\end{eqnarray}
which is reciprocal to the electric field $F_{rt}$, divided
additionally by $r^2$, consequently this gradient behaves,
at the center, inversely to $S(\mathcal{E})$, divided by $r^4$.\\
Thus for regular at the center solutions, for which the electric
field allows a series expansion beginning from $F_{rt}\simeq F_3
r^{3+i}+...,i=0,1,...,$ the order of the leading terms of the
related quantities is $S\approx r^{1+i}+...$, and
${\mathcal{L}}_\mathcal{F} \approx r^{-5-i}+...$, which determines
the rate at which ${\mathcal{L}}_\mathcal{F}$ approaches to
infinity, although $S$ approaches to zero.

\section{Regularity conditions for electrically charged SSS metrics
from linear integral solutions $Q(\mathcal{E})$}\label{RegCondN}

\subsection{Curvature invariants for the general metric function
$Q(\mathcal{E})$}

The general solution $Q(r)$ (\ref{EulerQ}) of the Einstein--NLE
equations derived via variations of parameters for a given electric
field $\mathcal{E}(r):=q_0F_{rt}(r)$ is given as
\begin{eqnarray}\label
{EulerQ1x}
Q=r^2-2m\,r-\frac{\Lambda_e}{3}\,{r}^{4}-\frac{2}{3}\,{r}^{4}\int \!
{\frac {\mathcal{E}
 }{{r}^{3}}}{dr} +\frac{2}{3}\,r\int \!\mathcal{E}
 {dr}.
\end{eqnarray}
Evaluating the curvature invariants $S$, $\Psi_2$, and $R$ one
arrives correspondingly at the following curvature invariants
\begin{eqnarray}\label{EulerQ2}&&
 {\it
S}=-\frac{1}{2}\,{\frac {{\mathcal{E}} \left( r \right)
}{{r}^{2}}},\nonumber\\&& {\Psi_2}=\frac{m}{r^3}+\frac{1}{6}\,{\frac
{\mathcal{E} \left( r \right) }{{r}^{2}}}-\frac{1}{3{r}^{3}}\,{
{\int \!\mathcal{E} \left( r \right) {dr}}{}}, \nonumber\\&& {\it
R}=4\,{\Lambda_e}+2\,{\frac {\mathcal{E} \left( r \right) }{{r}
^{2}}} +8\,\int \!{\frac {\mathcal{E} \left( r
 \right) }{{r}^{3}}}{dr}.
 \end{eqnarray}
As we pointed out before, the constant $m$ corresponds to the
``Schwarzschild mass''; for regular metrics it has to be equated to
zero.\\

The integrals can be considered as indefinite ones, or as definite
integrals in the limits from $0$ to $r$. Therefore the curvature
invariants of massless SSS electrically charged NLE metrics can be
given as
\begin{eqnarray}\label{EulerQ2DEF}&&
 {\it
S}=-\frac{1}{2}\,{\frac {{\mathcal{E}} \left( r \right) }{{r}^{2}}},
\nonumber\\&& {\Psi_2}=\frac{1}{6}\,{\frac {\mathcal{E} \left( r
\right) }{{r}^{2}}}-\frac{1}{3{r}^{3}}\,{{\int_0}^r \!\mathcal{E}
\left( r \right) {dr}}, \nonumber\\&& {\it
R}=4\,{\Lambda_e}+2\,{\frac {\mathcal{E} \left( r \right) }{{r}
^{2}}} +8\,{\int_0}^r \!{\frac {\mathcal{E} \left( r
 \right) }{{r}^{3}}}{dr}.
\end{eqnarray}

\subsection{Necessary regularity conditions}\label{NecessaryN}

The regularity of SSS electric NLE solutions in its $Q(\mathcal{E})$
representation  is achieved by establishing the regularity
conditions for the Riemannian curvature invariants
$\{\Psi_2(r),S(r),R(r)\}$, (\ref{EulerQ2DEF}), at the center.\\

The relation between $S$ and $\mathcal{E}$, (\ref{EulerQ2DEF}),
which can be written as
\begin{eqnarray}\mathcal{E}(r)=-2r^2\,S(r),
\end{eqnarray}
can be thought of as the series expansion of $\mathcal{E}$ near the
center with leading term proportional to $r^2$, with regular (near
the origin) invariant $S(r)$, equal to $S(0)$, i.e.,
\begin{eqnarray}\mathcal{E}(r)\simeq-2r^2\,S(0)+...,\lim_{r\rightarrow
0} S(r)=S(0),\,\text{ as } \,r\rightarrow 0,
\end{eqnarray}
and consequently at the center $\mathcal{E}(0)=0$; from the above
relation one gets: $\dot\mathcal{E}\simeq-4r\,S(0)$,\,
$\ddot\mathcal{E}\simeq-4\,S(0)$, therefore
\begin{eqnarray}\label{EQSE}
 \lim_{{r}\rightarrow
0}\{\mathcal{E},\dot{\mathcal{E}},\ddot{\mathcal{E}}\}=\{0,0,-4\,S(0)\}.
\end{eqnarray} Moreover, in this $r^2$--leading term approximation of
$\mathcal{E}$,  the evaluation of the invariant scalar curvature
leads to
\begin{eqnarray}
R(r)\simeq 4(\Lambda_e-S(r))-16 S(0)\,\ln(r),
\end{eqnarray}
hence to avoid the infinite logarithmical singularity at the center,
$S(0)$ ought to vanish, $S(0)=0$. This is a quite remarkable result,
the traceless Ricci tensor eigenvalue for regular electric SSS--NLE
solutions has to vanish at the center:
\begin{eqnarray}&&\label{Sequals0}
S(0)=0\Rightarrow \lim_{{r}\rightarrow 0}S(r)=0 \nonumber\\&&
\Rightarrow \lim_{r\rightarrow 0}\mathcal{E}=0, \lim_{{r}\rightarrow
0}\{ \mathcal{E},\dot{\mathcal{E}},\ddot{\mathcal{E}}\}=\{0,0,0\}.
\end{eqnarray}
Notice that in the $r^2$--leading term approximation of
$\mathcal{E}$ the Weyl invariant $\Psi_2$  leads to
\begin{eqnarray}\Psi_2(r)\simeq -\frac{1}{9}\,S(0)\Rightarrow \Psi_2(0)=0. \end{eqnarray}
On the other hand, the Lagrangian from (\ref{EulerQL}), in the
$r^2$--leading term expansion of $\mathcal{E}$, becomes
\begin{eqnarray}&&\label{EulerQL2}
\,\mathcal{L}(r)\simeq \mathcal{L}(0)- 2\,\int \!{\frac {S(0) }{{
r}}{dr}}-S(0)\nonumber\\&&=\mathcal{L}(0)-(2\ln(r)+1)S(0),\,
\lim_{{r}\rightarrow 0} \mathcal{L}(r)=\mathcal{L}(0),
\end{eqnarray}
where $\mathcal{L}(0)$ is a constant, associated to the local
energy density $\mu(0)$ defined in the weak and dominant energy conditions.\\
Therefore, the quadratic curvature invariants are regular at the
center if \begin{eqnarray}
 \lim_{{r}\rightarrow 0}\{
\mathcal{E},\dot{\mathcal{E}},\ddot{\mathcal{E}}\}=\{0,0,0\}.
\end{eqnarray}

Another alternative way of establishing the behavior of $\Psi_2$ and
$R$ at the center is by means of their integral representation
(\ref{EulerQ2DEF}). The regularity at the center of $\Psi_2$ leads
to
\begin{eqnarray}&&
\lim_{r \rightarrow 0}({\Psi_2}+\frac{1}{3}\,S)=-\lim_{r \rightarrow
0}(\frac{1}{3{r}^{3}}\,{{\int_0}^r \!\mathcal{E} \left( r \right)
{dr}})\nonumber\\&& \simeq -\frac{1}{3} FQ' \nonumber\\&&\Rightarrow
\,\lim_{r \rightarrow 0}{{\int_0}^r \!\mathcal{E} \left( r \right)
{dr}}\simeq FQ'\,r^3,
\end{eqnarray}
which, by differentiation or via the mean value theorem (with a
different coefficient of proportionality), near the center gives
rise to
\begin{eqnarray}\label{EexpFQ}\mathcal{E} \left( r \right)\simeq3 FQ'\,r^2,
 \text{as}\,\,r\rightarrow 0,
 \end{eqnarray}
which can be thought of from the series expansion point of view. The
substitution of this $\mathcal{E}$ into the scalar $R$ from
(\ref{EulerQ2DEF}) yields
\begin{eqnarray}&&
R(r)\simeq 4(\Lambda_e-S(r))+24 FQ'\ln(r),
\end{eqnarray} which is finite if $FQ'=0$, which in turn
implies, from (\ref{EexpFQ}), that $\mathcal{E}\rightarrow 0,\,$ hence
\begin{eqnarray}\label{EriesE}&&
FQ'=0,\,\mathcal{E}\rightarrow
0,\,{\mathcal{E}}\simeq\mathcal{E}_3r^3+...,
\end{eqnarray}
consequently
\begin{eqnarray}&&
 \lim_{r \rightarrow 0}S\rightarrow 0,\,\lim_{r
\rightarrow 0}R=4\Lambda_e,\,\lim_{r \rightarrow
0}{\Psi_2}\rightarrow 0.
\end{eqnarray}
Thus, the regularity of $S$, $\Psi_2$, and $R$ at the center takes
place for the series expansion for $\mathcal{E}$ (\ref{EriesE}). For
the series ${\mathcal{E}}\simeq\mathcal{E}_3r^3+...$,
\begin{eqnarray}&&
\lim_{r \rightarrow 0}(\frac{1}{3{r}^{3}}\,{{\int_0}^r \!\mathcal{E}
\left( r \right) {dr}})\rightarrow 0,\, \lim_{r \rightarrow
0}{{\int_0}^r \!\mathcal{E} \left( r \right) {dr}}\rightarrow 0,
\end{eqnarray}
are well behaved at the origin.

\subsection{Sufficient regularity conditions}\label{SufficientN}

On the other hand, let us assume for a moment that the above
conditions are the necessary ones but not the sufficient ones for
solution's regularity at the center; in the previous subsection from
the regularity requirements imposed on the curvature invariants of a
SSS solution  were established conditions (\ref{RegCOND}) for
regularity at the center.\\
To show that the conditions (\ref{RegCOND}) are  sufficient
conditions for regularity, we shall use the massless solution
representation (\ref{EulerQ1x}) of the metric function $Q$, which
fulfills Einstein-NLE equations in the form
(\ref{Einstein1Fyt2xcF}). This function and its first and second
derivatives are
\begin{eqnarray}\label{EulerQlM}&&
Q=r^2-\frac{\Lambda_e}{3}\,{r}^{4}-\frac{2}{3}\,{r}^{4}{\int_0}^r \!
{\frac {\mathcal{E}
 }{{r}^{3}}}{dr} +\frac{2}{3}\,r {\int_0}^r \!\mathcal{E}
 {dr},\nonumber\\&&
\dot{Q}=2r-\frac{4\Lambda_e}{3}\,{r}^{3}-\frac{8}{3}\,{r}^{3}{\int_0}^r
\! {\frac {\mathcal{E} }{{r}^{3}}}{dr} +\frac{2}{3}\,{\int_0}^r
\!\mathcal{E}
 {dr},\nonumber\\&&
\ddot{Q}=2-\,4\,\Lambda_e\,{r}^{2}-{8}\,{r}^{2}{\int_0}^r \!
{\frac{\mathcal{E}}{r^3}\,{dr}}-2\mathcal{E},
\end{eqnarray}
Evaluating them at the center as $r\rightarrow 0$, taking into
account that ${\int_0}^0 f(x)dx=0$ together with the finite
character of the integral ${\int_0}^r \!
{{\mathcal{E}}/{r^3}\,{dr}}$, in the series expansion
$\mathcal{E}\simeq \mathcal{E}_3r^3+...$, $\mathcal{E}_2=0$, one
gets
\begin{eqnarray}\label{EulerQlMx}&&
Q(0)=0,\,\,\,\dot{Q}(0)=0,\nonumber\\&&
\ddot{Q}(0)=2-2\mathcal{E}(0).
\end{eqnarray}
In order to have $\ddot{Q}(0)=2$, one has to set $\mathcal{E}(0)=q_0
F_{rt}(0)=0$, which is compatible with $\mathcal{E}\simeq
\mathcal{E}_3r^3+...$ at zero, or, $\lim_{{r}\rightarrow
0}\{\mathcal{E},\dot{\mathcal{E}},\ddot{\mathcal{E}}\}=\{0,0,0\}.$
Therefore, we established that the conditions $\lim_{{r}\rightarrow
0}\{{Q},\dot{{Q}},\ddot{{Q}}\}=\{0,0,2\}$ have to hold too.

Recall that the $r^2$--term has to be present in the metric function
$Q=r^2+...$ to guaranty the Lorentzian character,
$\text{diag}(1,1,1,-1)$, of the metric line element.

These results can be gathered in the form of a theorem:\\
The Einstein--NLE theory allows for electric SSS solutions with
regular curvature invariants at the origin $\lim_{{r}\rightarrow
0}\{S,\Psi_2,R\}= \{0,0,(0,4\,\Lambda+4\mathcal{L}(0)\}$ if and only
if the following regularity conditions at the center hold:
\begin{eqnarray}\label{REGCON}&&
\lim_{r\rightarrow
0}\{\mathcal{E},\dot{\mathcal{E}},\ddot{\mathcal{E}}\}=\{0,0,0\},
\nonumber\\&& \lim_{{r}\rightarrow
0}\{Q,\dot{Q},\ddot{Q}\}=\{0,0,2\}.
\end{eqnarray}
As a consequences of these conditions, one gets
\begin{eqnarray}&&
\lim_{{r}\rightarrow 0}\{{F_{{r}t}},\,{\mathcal{F}},
{\mathcal{F}}\mathcal{L}_{\mathcal{F}},{\mathcal{L}}\}=\{0,0,0,\mathcal{L}(0)\}.
\end{eqnarray}
Regular nonlinear electrodynamics electric static spherically
symmetric solutions approach to the flat or conformally flat dS--AdS
(regular) spacetimes at the center.\\

From the remaining independent electric field equation, one
establishes
\begin{eqnarray}\label{FIELD}&&
\mathcal{L}_{\mathcal{F}}=-\frac{q_0}{r^2F_{rt}}\Rightarrow
\lim_{{r}\rightarrow 0}\mathcal{L}_{\mathcal{F}}\rightarrow\infty,
\end{eqnarray}
at the center. Notice that equation
$S=-{\mathcal{F}}\,\mathcal{L}_{\mathcal{F}}$, yields
$$\lim_{{r}\rightarrow 0}S=-\lim_{{r}\rightarrow
0}\left({\mathcal{F}}\,\mathcal{L}_{\mathcal{F}}\right)=0,$$
although the limit of each term behaves opposite one to another; $
\lim_{{r}\rightarrow 0}\,{\mathcal{F}}=0$ and $\lim_{{r}\rightarrow
0}\,\mathcal{L}_{\mathcal{F}}=\infty$; the limit of $S$ has to be
considered as a whole. A similar situation happens in the classical
limit  $\lim_{r\rightarrow 0}\,\frac{\sin(r)}{r}$ which, as an
entity, gives 1, $\lim_{r\rightarrow 0}\,\frac{\sin(r)}{r}=1$, but
with term by term limits: $\lim_{r\rightarrow 0}\,{\sin(r)}=0$, and
$\lim_{r\rightarrow 0}\,\frac{1}{r}\rightarrow\infty$.

\section{ SSS solutions for a polynomial $\mathcal{E}=q_0\,F_{rt}$ field }\label{seriesEXP}

Let us consider the class of solutions determined by the field
component $F_{rt}$ of the following polynomial form
\begin{eqnarray}\label{FEuler1}
\mathcal{E}=q_0\,F_{rt}=f_0+ f_1\,r +f_2\,r^2+f_3\,r^3+r^4\sum_0^n
p_i\,r^i,
\end{eqnarray}
named usually ``interpolation polynomial'' in the approximation
method of integration, where $ p_i$ are constants that may assume
even the value $0$. One may consider (\ref{FEuler1}) also as a
Taylor series expansion up to certain order $n+4$; for $p_i\equiv
0$, $q_0\,F_{rt}$ from (\ref{FEuler1}) becomes a cubic solution or a
series expansion up to the third order. By means of this polynomial
function we intent to answer the question: which are the powers
$r^k$ in $F_{rt}$ for which the curvature regularity fails?\\

The function Q(r), using the general solution (\ref{EulerQlM}) for
the given field $\mathcal{E}$ (\ref{FEuler1}), amounts to
\begin{eqnarray}\label{MetricforP}
&& Q(r)={r}^{2}+{f_0}\,{r}^{2}+{ f_1}\,{r}^{3}-\frac{1}{2}\,{
f_3}\,{r}^{5} -2\,m
r-\Lambda_e{r}^{4}/3\nonumber\\&&-\frac{2}{3}\,{f_2}\,{r}^{4}\ln
\left( r \right) -2\,r^6 \,\sum_0^n \frac{p_i}{(i+2)(i+5)} \,r^i.
\end{eqnarray}
Evaluating the curvature invariants (\ref{EulerQ2}) and the
Lagrangian $\mathcal{L}$ from (\ref{EulerQL}), with
$\Lambda_e=\Lambda+ \mathcal{{ L}}(0)$, one gets
\begin{eqnarray}
&& 2{S}=-\frac{f_0}{{r}^{2}}-\frac{f_1}{{r}}-f_2-f_3r-r^2\, \sum_0^n
p_i\,r^i ,\nonumber\\&&{
R}=4\,\Lambda_e-2\frac{f_0}{r^2}-6\frac{f_1}{r}+\frac{14}{3}f_2+8\,{
f_2}\,\ln \left( r \right) \nonumber\\&&+10\,f_3 r+2r^2\sum_0^n
\frac{i+6}{i+2}p_i\,r^i ,\nonumber\\&&{\Psi_2}={\frac
{m}{{r}^{3}}}+\frac{{ f_2}}{18}\,+\frac{{
f_3}}{12}\,\,r-\frac{1}{6}\,{\frac {{ f_0}
}{{r}^{2}}}+\frac{r^2}{6}\sum_0^n\frac{i+3}{i+5} p_i\,r^i,
\nonumber\\&& \mathcal{{ L}}= \mathcal{{
L}}(0)+\frac{5}{3}f_2+2\,{f_2}\,\ln \left( r \right)-{\frac
{{f_1}}{r}}+3 \,{f_3}\,r \nonumber\\&&+r^2\sum_0^n\frac{i+4}{i+2}
p_i\,r^i.
\end{eqnarray}
\marginpar{$\mathcal{ L}(0)$ OK  line1273} This set of functions
characterizes completely the SSS metric for a polynomial $F_{rt}$
function of the Einstein--electrically charged NLE: this metric is
endowed with mass, cosmological constant and a set of electric
parameters associated to the nonlinearity of the electrodynamics. In
general, this solution is singular at the center due to the presence
of logarithmical functions, and terms proportional to $1/r$, and
$1/r^2$ in curvature functions.

\subsection{Regularity}

One may consider the above expression (\ref{MetricforP}) for $Q$ as
the solution for the series expansion near the origin of a well
behaved field function $F_{rt}$. To avoid infinities in the
curvature quantities, it is clear that $f_2$ ought to vanish in
order to drop out the terms with $\ln(r)$, the mass term with $m$,
as well as $f_0$, have to vanish in order to regularize $\Psi_2$,
moreover $f_1$ has to be zero for the regularity of $S$ and
$\mathcal{L}$ too. Therefore, this NLE solution becomes a regular
one if
\begin{eqnarray}\{f_0,f_1,f_2\}=\{0,0,0\}
\equiv \lim_{{r}\rightarrow
0}\{\mathcal{E},\dot{\mathcal{E}},\ddot{\mathcal{E}}\}=\{0,0,0\}.
\end{eqnarray}
Hence the field $F_{rt}$ becomes a polynomial of power equal or
greater than $3$. This NLE regular solution is given by
\begin{eqnarray}&&{ q_0F_{rt}}= {f_3}\,{r}^{3}+\,r^4
\,\sum_0^n\,p_i\,r^i,\,\nonumber\\&&
 Q(r)={r}^{2}-\Lambda_e\, r^4/3-\frac{1}{2}\,{
f_3}\,{r}^{5} \nonumber\\&&-2\,r^6 \,\sum_0^n \frac{p_i}{(i+2)(i+5)}
\,r^i,
\end{eqnarray}
with invariant functions
\begin{eqnarray}&&
 2{S}=-f_3\,r-r^2\, \sum_0^n p_i\,r^i ,\nonumber\\&&
{ R}= 4\Lambda_e+10\,f_3 r+2r^2\sum_0^n \frac{i+6}{i+2}p_i\,r^i
,\nonumber\\&&{\Psi_2}=\frac{{
f_3}}{12}\,r+\frac{r^2}{6}\sum_0^n\frac{i+3}{i+5} p_i\,r^i,
\nonumber\\&& \mathcal{{ L}}=\mathcal{{ L}}(0)+3
\,{f_3}\,r+r^2\sum_0^n\frac{i+4}{i+2} p_i\,r^i.
\end{eqnarray}
Incidentally, from this series expansion point of view, one can
establish the expansion behavior of the gradient
$\mathcal{L}_\mathcal{F}=-q_0/F_{rt}/r^2$, appearing in
$S=-\mathcal{F}\mathcal{L}_\mathcal{F}$. For an expansion
$F_{rt}\simeq r^k +O(r^{k+1})$, at the center, one gets
\begin{eqnarray}&&
F_{rt}\simeq r^k\Rightarrow\mathcal{L}_\mathcal{F}\simeq
r^{-k-2},S\simeq r^{k+2}.
\end{eqnarray}
For regular solutions, with an electric field $F_{rt}$  with
$r^{3+i}$--leading term expansion, $ i=0,1...$, one has
\begin{eqnarray}&&
F_{rt}\simeq r^{3+i}\Rightarrow\mathcal{L}_\mathcal{F}\simeq
r^{-5-i},S\simeq r^{1+i}.
\end{eqnarray}
 At the center, the behavior
of the metric function $Q(r)$ and the electric field $\mathcal{
E}=q_0F_{rt}$ and their first and second derivatives is
\begin{eqnarray*}&&\lim_{r\rightarrow 0}\{Q,\dot Q,\ddot Q\}=\{0,0,2\}
,\nonumber\\&& \lim_{r\rightarrow 0}\{\mathcal{E},\dot\mathcal{
E},\ddot \mathcal{ E}\}=\{q_0f_0,q_0f_1,q_0f_2\}=\{0,0,0\}.
\end{eqnarray*}
Thus, at the center $r=0$, the gravitational field $g_{\mu\nu}$
reduces to the corresponding one of the flat or (A)dS--like metric (
if a $\Lambda_e$--term is present), which is in fact the expected
flat--(A)dS spacetime limit
\begin{eqnarray}&&
Q(r)=r^{2}-\Lambda_e{r}^{4}/3 ,\,\,\,g_{tt}=-
Q(r)/{r}^{2},\nonumber\\&& { R}=4\Lambda_e,\,{\Psi_2}(0)=0,\,
S(0)=0.
\end{eqnarray}

\section{ Maxwell asymptotic at infinity}

In the Maxwell theory the only SSS electric solution is
Reissner--Nordstr\"om (RN) spacetime determined by a single
component $F_{rt}=-q_0/r^2$, hence
$\mathcal{F}=-q_0^2/(2r^4)=\mathcal{L}$ in the whole spacetime, with
all its curvature invariants singular at the origin.

The Einstein--NLE electric solutions allowing for a Maxwell weak
field limit (small) $\mathcal{F}$, $\{\mathcal{L}\rightarrow
\mathcal{F}, \,\mathcal{L}_{\mathcal{F}}\rightarrow 1\}$, reduce to
the Reissner--Nordstr\"om (RN) spacetime.\\ Let us analyze in detail
the content of the equation (\ref{electric}); at spatial infinity
${r}\rightarrow \infty$, the Einstein--Maxwell asymptotic solution
should be the Reissner--Norstr\"om solution, determined by
 $\mathcal{F}=-F_{{r}t}^2/2$, as $ F_{{r}t}$ approaches to $-{q_0}/{{r}^2}$,
  for short $ F_{{r}t}\rightarrow-{q_0}/{{r}^2}$,
  which substituted in the equation
 (\ref{electric}) gives
$$ {r}^2\,{F_{{r}t}}{\mathcal{L}}_{\mathcal{F}}=-q_0
 \Rightarrow{{\mathcal{L}}_{\mathcal{F}}\rightarrow1}
 \Rightarrow{{\mathcal{L}}\rightarrow{\mathcal{F}}},$$
i.e., (\ref{electric}) gives rise to the correct Maxwell linear
condition. On the other hand, assuming the limiting Maxwell
character of the electrodynamics,
${\mathcal{L}}_{\mathcal{F}}\rightarrow 1$, from (\ref{electric})
one gets
$${r}^2\,{F_{{r}t}}{\mathcal{L}}_{\mathcal{F}}=-q_0
\Rightarrow{F_{{r}t}}\rightarrow-q_0/{r}^2,$$
i.e., one arrives at the electric field for a central charge in the
Maxwell theory. Therefore, there is a subclass of electric SSS
metrics that approaches to the Maxwell Reissner--Nordstr\"om
electric solution at infinity. These solutions exhibit at
 infinity the Maxwell asymptotic
($
{\mathcal{L}}\rightarrow\mathcal{F},{\mathcal{L}}_{\mathcal{F}}\rightarrow
1\,$ for weak field $\mathcal{F}$), together with the Maxwell field
limits: ${F_{{r}t}}\rightarrow-q_0/{r}^2,\,{\mathcal{F}}\rightarrow
-q_0^2/(2{r}^4)$.\\
Incidentally, as far as to electrically charged SSS solutions to
Einstein--NLE equations allowing for Maxwell asymptotic at the
center, there is a ``no--go'' theorem by Bronnikov
\cite{BronnikovPRD2002} that asserts the charge has to be zero,
i.e., there is no such a kind of solutions. The demonstration is
quite elementary: assume that the Maxwell asymptotic $
 (\mathcal{L}\rightarrow 0,\mathcal{L}_{\mathcal{F}}\rightarrow 1
  \,\,\text{as} \,\,\mathcal{F}\rightarrow 0)$ holds at the center,
using these conditions in the electric field equation
(\ref{electric}), $r^2\,\mathcal{L}_\mathcal{F}\,F_{rt}=-q_0$, one
arrives at
 \begin{eqnarray}\label{FieldEqresLim}&&
\lim_{r\rightarrow 0}\left(\,r^2\mathcal{L}_\mathcal{F}\,F_{rt}
\right)=-q_0,\nonumber\\&& \lim{\,r^2}\times\lim
\left(\,\mathcal{L}_{\mathcal{F}}F_{rt}\right)=-{q_0}, {\text{
as}}\,\, r\rightarrow 0,\nonumber\\&& \Rightarrow 0\,\times \lim
\left(\,\mathcal{L}_{\mathcal{F}}F_{rt}\right)=-q_0\Rightarrow
q_0=0,
\end{eqnarray}
thus one gets that the charge is zero. Consequently, the electric
field equation
 (\ref{electric}), for $q_0=0$, leads to
\begin{eqnarray}\label{FieldEqresaN}
r^2\,L_F \,F_{tr}=0\Rightarrow  F_{tr}=0\Rightarrow \mathcal{F}=0,
\,{\text{for all $r$ }}.
\end{eqnarray}
The absence of the electric fields, for a vanishing electric charge
constant $q_0$, in the whole SSS spacetime is an expected and
non--surprising at all result pointing on the no existence of NLE,
for $\mathcal{L}(\mathcal{F})$, under the Maxwell asymptotic for a
SSS spacetime with a regular center; in such a case the SSS metric
becomes the flat or vacuum with a cosmological term spacetime.

\section{Weak and dominant energy
conditions for electrically charged NLE SSS metrics}\label{WECDC}

In this section the relation between the value at the origin of the
Lagrangian $\mathcal{L}$ and the energy density $\mu$ is
established, namely $\mathcal{L}(0)=\mu(0)$. Following Hawking and
Ellis (HE) \cite{Hawking73}, in a real universe  it is practically
impossible to establish all the energy-momentum (EM) tensors,
contributing to the Einstein theory. Nevertheless, one can impose
certain physically acceptable conditions--inequalities--on the EM
tensor, see HE \cite{Hawking73}, \S 4.3, where one reads ``In many
circumstances these (conditions) are sufficient to  prove the
occurrence of singularities,
independent of the exact form of the energy--momentum tensor.''\\
Weak energy condition (WEC): for any timelike vector $V^a$, $V_a
V^a=-1$, the energy--momentum (EM) tensor $T_{ab}$ obeys the
inequality $T_{ab}V^aV^b\geq 0$, which means that local energy
density $T_{ab}V^aV^b\geq 0$ as measured by any observer with
timelike vector $V^a$ is a non--negative quantity. The component of
the $T_{ab}$ are defined with respect to an orthonormal basis
$\{E_1,E_2,E_2,E_4\}$, where $E_4$ is timelike unit vector, and the
HE type I EM tensor with respect to this basis is diagonal. On this
respect in \cite{KramerStephani03}\,\S\,5.3 one reads: ``For
energy--momentum tensors of the Segre--Pleba\'nski type
$[111,1]\simeq[S_1-S_2-S_3-T]_{(1111)}$ (and its degeneracies)
$T_{ab}$ can be diagonalized $T_{ab}=(p_1,p_2,p_3,\mu)$'' and their
components satisfy the weak energy conditions if:
\begin{eqnarray}\label{WECX}
\mu=T^{44} \geq 0,\,\mu+p_k \geq 0,\,k=1,2,3,
\end{eqnarray}
or a more restrictive dominant energy condition (DEC):
 for any timelike vector $W^a$,
the energy--momentum tensor $T_{ab}$ obeys the inequality
$T_{ab}W^aW^b\geq 0$ and the local energy flow vector $T_{ab}W^a$ is
a non--spacelike vector. For type I EM tensors the dominant energy
condition can be thought of as the WEC with the additional
requirement that the pressure should no exceed the energy density.
This condition leads to
\begin{eqnarray}\label{WECX}
\mu =T^{44}\geq 0,\,\mu+p_k \geq 0,\,\,\mu-p_k \geq 0,k=1,2,3,
\end{eqnarray}
which points on the fact that the energy dominates the other
components of $T_{ab}$
\begin{eqnarray}
T^{44}\geq |T^{ab}|, {\text {for each a,b.}}
\end{eqnarray}
This type, type I in HE wording, of energy--momentum tensors
\begin{equation}\label{hetypeI}
{T^{a}}_{b}=p_1 {\delta^a}_1{\delta^1}_b+p_2
{\delta^a}_2{\delta^2}_b+p_3 {\delta^a}_3{\delta^3}_b-\mu
{\delta^a}_4{\delta^4}_b,
\end{equation}
describes fluids, and its sub--types describe non--null
electromagnetic fields, perfect fluids, cosmological
$\Lambda$--terms, among others. In the case of nonlinear
electrodynamics for SSS metrics, the electrodynamics EM tensor, for
$\mathcal{L}=\mathcal{L}(\mathcal{F})$, amounts to
\begin{equation}\label{Hawkinga}
({T^{a}}_{b})={\text{diag}}(-\mathcal{L},-\mathcal{L},-(\mathcal{L}+2S)
,-(\mathcal{L}+2S)), S=-{\mathcal{F}} \,\mathcal{L}_{\mathcal{F}}.
\end{equation}
Therefore, comparing (\ref{hetypeI}) with (\ref{Hawkinga}), one gets
\begin{eqnarray}&&\label{Hawking1}
   P_{\theta}=P_{\phi}= -\mathcal{L},\,\,\,P_{r}= - \left(\mathcal{L}
+2S\right),\nonumber\\&& \mu(r)=\mathcal{L} -2{\mathcal{F}}
\,\mathcal{L}_{\mathcal{F}} =\mathcal{L}+2S,\,\mu(r)+P_r(r)=0.
 \end{eqnarray}
In accordance with the weak--dominant energy conditions, one arrives
for NLE at the following relations:
\begin{eqnarray}&&\label{WDENC}
\mu=\mathcal{L}-2\mathcal{F}\mathcal{L}_\mathcal{F}\geq 0,\,\,
\Rightarrow
\mathcal{L}(\mathcal{F})\geq2\mathcal{F}\mathcal{L}_\mathcal{F},\nonumber\\&&
\lim_{r\rightarrow0}\mu=\lim_{r\rightarrow
0}\mathcal{L}+2\lim_{r\rightarrow 0}S=\mathcal{L}(0),
\lim_{r\rightarrow 0}S=S(0)=0, \nonumber\\&&
\Rightarrow\mu(0)=\mathcal{L}(0)\geq 0,
\end{eqnarray}
\begin{eqnarray}&&\label{ag2bx}
\mu+P_{\phi}=-2\mathcal{F}\,\mathcal{L}_\mathcal{F}\geq 0;\text
{since}\, \mathcal{F}\leq 0,
{\text{then}}\,\,\mathcal{L}_\mathcal{F} \geq 0,
\nonumber\\&&\lim_{r\rightarrow 0} (\mu+P_{\phi})=\lim_{r\rightarrow
0}(-2F\,L_F)=2\lim_{r\rightarrow 0}S=0 ,
\end{eqnarray}
\begin{eqnarray}&&\label{DOMM}
\mu-P_{\phi}=2\mathcal{L}+2S,\,\lim_{\rightarrow
0}(\mu-P_{\phi})=2\mathcal{L}(0)\geq 0,\nonumber\\&&
\mu-P_r=2\mathcal{L}+2S,\,\lim_{r\rightarrow
0}(\mu-P_r)=2\mathcal{L}(0)\geq 0.
\end{eqnarray}
On the other hand, from the Einstein equations, one may derive the
conservation equation for the EM tensor ${T^{\mu\nu}}_{;\mu}=0$; the
Einstein--NLE SSS equations (\ref{EQET1}) can be written in the HE
fluid language as
\begin{eqnarray}\label{EQET1HE}
&&E^t_t;\frac{\dot{Q}}{{r}^{3}} -{ \frac {Q }{{r}^{4}}} -\frac
{1}{{r}^{2}}+\Lambda=-\mu(r),\nonumber\\&& E^\phi_\phi
:\frac{\ddot{Q}}{2\,{r}^{2}} -\frac{\dot{Q}}{{r}^{3}}
 +{ \frac {Q }{{r}^{4}}} +\Lambda =P_{\phi}.
\end{eqnarray}
Isolating the derivative $\dot{Q}$ from $E^t_t$ and substituting it
into the equation $E^\phi_\phi$ one arrives at the conservation
equation in the form
\begin{eqnarray}\label{EQET11}
\mu+\frac{r}{2}\frac{d\mu}{dr}=-\mathcal{L}=-P_{\phi},
\end{eqnarray}
which leads to
\begin{eqnarray}&&\label{ag2a}
0 \leq\mu+P_{\phi}=-\frac{r}{2} \frac{d\mu}{dr}\Rightarrow
\frac{d\mu}{dr}\leq 0.
\end{eqnarray}
Consequently, for physically reasonable NLE theories
$\mathcal{L}(\mathcal{F})$, the curve $\mu(r)(=
\mathcal{L}-2\mathcal{F}\,\mathcal{L}_\mathcal{F})\geq 0$ has to be
a decreasing function (it has a negative slop) from a maximal value
at $r=0$, $\mu(0)=\mathcal{L}(0)\geq 0$, which represents the
maximal value of the local energy density $\mu$ at the origin. This
condition  $\mu(0)=\mathcal{L}(0)\geq 0$ guarantees, taking into
account $\lim_{r\rightarrow 0} S(r)=0$, the fulfilment of all
remaining weak--dominant conditions.\\ As far as the behavior at the
center of the EM tensor (\ref{Hawkinga}) is concerned one gets
\begin{equation}\label{Hawkingab}
 \lim_{r\rightarrow 0}({T^{a}}_{b})
={\text{diag}}(-\mathcal{L}(0),-\mathcal{L}(0),-\mathcal{L}(0),\mathcal{L}(0)),
\end{equation}
which is a cosmological $\Lambda$--term $[(111,1)]\simeq [4T]_{(1)}$
tensor type. Notice that the solution for the Einstein equations
(\ref{EQET1HE}), compatible with the EM tensor (\ref{Hawkingab}) as
$r\rightarrow 0$, is given by the familiar Kottler--like metric
function
\begin{eqnarray} \label{ADSl}
Q=r^2-\frac{\Lambda_e}{3}
r^4\equiv{r^2-\frac{\Lambda+\mathcal{L}(0)}{3} r^4}.
\end{eqnarray}
Moreover, the relation energy--momentum trace--scalar curvature
\begin{equation}\label{CruRwec}
R(r)=4\Lambda+4\mathcal{L}(r)+4{
S}(r)=4\Lambda+2\mathcal{L}(r)+2\mu(r)
\end{equation} at
the center leads to
\begin{eqnarray*}\label{CruRwec}
R(0)=4\Lambda+4\mathcal{L}(0),
\end{eqnarray*}
when weak--dominant energy conditions hold.

\section{Linear superposition of SSS electric solutions of the
Einstein--NLE equations}\label{LinearQ}

Due to the linearity of the equations (\ref{Einstein1Fyt2xcF}) one
can decompose the metric function $Q(r)=Q_K+ Q(\mathcal{E})$, where
$Q(\mathcal{E})$, responding to the field $\mathcal{E}$, fulfils
\begin{equation}\label{Einstein1op}
D_fQ(\mathcal{E})= -2r^2 \mathcal{E} \left( {r} \right),\,\,
D_LQ(\mathcal{E})= -2r^4\mathcal{L}({r}),
\end{equation}
thus, for any given function $Q(\mathcal{E})$ one determines a
Lagrangian function $\mathcal{L}(\mathcal{E})$ and the corresponding
field component $q_0\,{F}_{rt}=:\mathcal{E}$.
 On the other hand,
$Q_K$ is the Kottler-like function, responding to the vacuum--
$\Lambda$--term equations,
\begin{eqnarray}&&
D_fQ_K=-2r^2,D_LQ_K=-2r^4\Lambda_e,\nonumber\\&&
Q_K=r^2-2mr-\frac{\Lambda_e}{3}r^4,
\,\Lambda_e=\Lambda+\mathcal{L}(0).
\end{eqnarray}
Since these equations (\ref{Einstein1op}) depend linearly on the
structural function $Q(\mathcal{E})$ and its derivatives, a linear
superposition, with constants $ C_i$, of metric functions
$Q(\mathcal{E}_i)$, , where
$\mathcal{E}_i:={q_0}_{i}\,{{F}_{rt_{i}}}$, no sum in $i$, is in
order, $Q_K+ \sum_i C_i Q(\mathcal{E}_i)$. The arbitrarily given
structural functions $Q_{i}:=Q(\mathcal{E}_i)$, $i=1,...,n$,
equipped with its own set of parameters $
p_i=\{{p_i}_j,\,j=1,...,s\}$, yields, via their substitution into
equations (\ref{Einstein1op}), a linear superposition of the
Lagrangian functions $\mathcal{L}_i:=\mathcal{L}(\mathcal{E}_i)$ and
the electric fields $\mathcal{E}_i$; the electric field equation
\begin{eqnarray}\label{EQfieldE}
 r^2\frac{d\mathcal{L}_i}{dr}- \frac{d\mathcal{E}_i}{dr}=0 ,
 \end{eqnarray}
are fulfilled identically. This superposition property can be formulated as:\\
  Theorem: In the framework of static spherically symmetric
metrics coupled to electric electrodynamics (linear and non-linear)
and a cosmological constant, any given functions
$Q_i(r):=Q(\mathcal{E}_i)$,
$\mathcal{E}_i:={q_0}_{i}\,{{F}_{rt}}_{i}$, no sum in $i$, gives
rise, via the Einstein--NLE equations, to a pair of NLE
electromagnetic functions $\{\mathcal{L}_i,\mathcal{E}_i\} $. Any
linear superposition of functions $Q_i(r)$, $i=1,...,n$, yields to a
linear superposition of the corresponding Lagrangian functions
$\mathcal{L}_i$ and the electromagnetic field functions
$\mathcal{E}_i$, fixed $i$; the Einstein--NLE electric field
equation are fulfilled identically: schematically
\begin{eqnarray}&&\,
{E^\mu}_{\nu}(\{Q_{i};\mathcal{L}(Q_{i}),\mathcal{E}(Q_{i})\})=0,\nonumber\\&&
\mathcal{Q}_T=Q_K+\sum_i
C_iQ_{i}(r),\nonumber\\&&{\mathcal{L}_T=\mathcal{L}(0)+\sum_i C_i
\mathcal{L}_{i}(r),\,\,\mathcal{E}_T=\sum_i C_i \mathcal{E}_{i}(r)},
\nonumber\\&&\, {E^\mu}_{\nu}(\{\,
\mathcal{Q}_T;\mathcal{L}_T,\mathcal{E}_T\})=0,
\end{eqnarray}
where $Q_K$ is the vacuum plus $\Lambda_e$--term Kottler solution.
The resulting solution will be characterized by the curvature
invariants $S_T=\sum_i C_i S_i$, $\Psi_{2T}=\sum_i C_i \Psi_{2i}$,
and $R_T=4\Lambda_e+\sum_i C_i
R_i$.\\
Notice that, in this formulation, the necessary metric term  $r^2$
to guarantee the signature (1,1,1,-1), the optional mass $-2m\,r$,
and the $\Lambda_e$--term appear only once through $Q_K$ in the
superposed functions.

The spacetimes allowing for regular curvature invariants at the
center can be thought of as immersed in a dS--AdS or in a flat
universe.

\subsection{Linear superposition for the general $Q(\mathcal{E})$
solution}\label{LinearQE}

Another perspective on linear superpositions of SSS solutions can be
achieved from the metric function $Q(\mathcal{E})$, (\ref{EulerQ}).
The $Q(\mathcal{E})$ function can be given in the form
\begin{eqnarray}\label{EulerQ1a}&&
{Q(r)}=Q_{K}-\frac{2}{3}\,{r}^{4}\int \! {\frac {\mathcal{E} \left(
r \right) }{{r}^{3}}}{dr} +\frac{2}{3}\,r\int \!\mathcal{E}\left( r
\right) {dr},\nonumber\\&&\,Q_{K}:={r}^{2}-2{\it
m}\,r-\frac{\Lambda_e}{3}\,{r}^{4},\, \mathcal{E}:=q_e F_{tr}(r).
\end{eqnarray}
where the subscript $K$ in the function $Q_K$ stands for
Kottler--like metric function \cite{Kottler}. For a linear
superposition of solutions, it is enough to accomplish the
superposition of any number of electric field functions
$\mathcal{E}_i,\, i=1,...,n$, each of them equipped with its own set
of parameters ${p_i}_k,\,k=1,...,s$, if any. The integrals will
generate new sets of functions endowed with the set of constants
$\{{p_{i}}_k,\, i=1,...,n,\,k=1,...,s\}$. Consequently, symbolically
one has:\\ $ \mathcal{E}= \sum_i C_i\mathcal{E}_i$,
\,$\mathcal{E}_j:={q_e}_j{F_{rt}}_j$, no sum in $j$,
\begin{eqnarray}\label{EulerQ1b}&&
Q(r)= Q_K+{Q(\mathcal{E})},\,\,\,Q_{K}:={r}^{2}-2{\it
m}\,r-\frac{\Lambda_e}{3}\,{r}^{4},
\nonumber\\&&{Q(\mathcal{E})}:=Q\left({\sum_i C_i
\mathcal{E}_i}\right)=\sum_i C_i Q\left({ \mathcal{E}_i}\right),
\nonumber\\&& Q({\mathcal{E}}_i):=\frac{2}{3}\,r\int{
\mathcal{E}_i}{dr}-\frac{2}{3}\,{r}^{4}\int \! {\frac {
\mathcal{E}_i}{{r}^{3}}}{dr} , \nonumber\\&&
\end{eqnarray}
with the associated Lagrangian
\begin{eqnarray}\label{EulerQ2a}
\mathcal{L}=\mathcal{L}(0)+\sum_iC_i \mathcal{L}({\mathcal{E}}_i)
,\,\mathcal{L}({\mathcal{E}}_i)={\frac {{\mathcal{E}}_i
}{{r}^{2}}+2\int \!{\frac {{\mathcal{E}}_i \left( r \right) }{{
r}^{3}}}{dr}}.
\end{eqnarray}

 \section{ Integrals of the SSS Euler equations associated to
$\Psi_2$ and $\mathcal{L}$ }\label{EulerCurInv}

The Einstein--NLE field equations for SSS metrics can be analyzed
from the point of view of Euler equations. In fact, beside the
scalar curvature $R$, (\ref{curvaR}), these equations can be written
as
\begin{eqnarray}\label{Einstein1Fyt2}&&
r^2{\ddot{Q}}  -4 r\,{\dot{Q}}+4\,{Q}= -2r^2-2r^2{ q_0}\, F_{{r}t}
\left( {r} \right),\nonumber\\&&r^2{\ddot{Q}} -2\,r {\dot{Q}}
+2\,{Q}= -2r^4\Lambda-2r^4\mathcal{L}({r}) ,\text{sign minus
ok}\nonumber\\&& {{r}}^{2}\ddot{Q} -6\,{r}\, \dot{Q} +12 \,Q
=2\,{{r}}^{2}-12\,{r}^4\,\Psi_2(r),
 \end{eqnarray}
and can be integrated through the variation of parameters of the
homogeneous solutions to these Euler equations.
 \subsection{
Solutions in terms of linear integrals of the Lagrangian function
$\mathcal{L}$}

For instance, one may proceed with the Euler equation for the
Lagrangian function and derive
\begin{eqnarray}\label{QspecLa}
Q_{\mathcal{L}}&&= Q_K+2\, r \int{r^2\,\mathcal{L} \,dr}-2\,r^2\int
{r\,\mathcal{ L}\,dr},
\end{eqnarray}
which is accompanied with the electric field
\begin{eqnarray}\label{QspecLaE}
q_0F_{rt}= r^2\mathcal{L}- 2\,\int r\,\mathcal{L}(r) dr=-2\,r^2\,S,
\end{eqnarray}
and characterized by the Weyl curvature component
\begin{eqnarray}\label{QspecLaEW}
\Psi_{2}= \frac{m}{r^3}-\frac{1}{r^3}\int
r^2\,\mathcal{L}dr+\frac{1}{3r^2} \,\int r\,\mathcal{L}
dr+\frac{1}{6}\mathcal{L},
\end{eqnarray}
and the scalar curvature $R$ fulfilling $ R=4\Lambda+4\mathcal{L}+4S
$, in this last relation and in the equations containing
(indefinite) integral of $\mathcal{L}$ the evaluation at zero does
not contribute with $\mathcal{L}(0)$, it is present only in
$\Lambda_e=\Lambda +\mathcal{L}(0).$

\subsection{ Solutions in terms of linear integrals of the Weyl
curvature component $\Psi_2$} If one were searching for a spacetime
with a specific behavior of the Weyl conformal tensor, then one
would integrate the third Euler equation with the $\Psi_2$--term;
the homogeneous equation for $\Psi_2$ possesses two independent
solutions $Q_1 =r^4$ and $Q_2=r^3$. Using the method of variations
of parameters, one gets
\begin{eqnarray}\label{QspecPSI}
Q_{\Psi_2}&&=r^2+ A_0r^4+B_0 \,r^3 \nonumber\\&&+ r^3\int{12\Psi_2
dr}-r^4\int{12\Psi_2\,/r\,dr}.
\end{eqnarray}
This metric function $Q_{\Psi_2}$ possesses the specific print of
the given Weyl function $\Psi_2(r)$, which is inherited to the
spacetime itself via the evaluation of the remaining functions $ q_0
F_{rt}\sim S$, $\mathcal{L}$, and $R$.\\
This approach can be considered as an alternative way to determine
solutions with particular properties. The case $q_0 F_{rt}\simeq S$
has been integrated previously and developed in detail here in
various paragraphs.

\subsection{Analyticity of the field functions}\label{Analyticity}

The description of physically relevant fields is expected to be done
by well--behaved (analytical) functions. Nevertheless, if one has in
mind the description of  $\mathcal{L}(\mathcal{F})$, in general, one
may find troubles in expressing $r=r(\mathcal{F})$, explicitly in
terms of $\mathcal{F}$, because of the possible appearance of
transcendent equations. In the case of ``regular at the origin and
Maxwell at infinity'' solutions, in general, the graph of
$F_{{r}t}({r})$ begins from zero in the origin, evolves (grows up or
decreases), reaches its maxima and minima, and again, at spatial
infinity approaches (from above or below) to zero; the existence of
a extremum, where $dF_{rt}/d{r}=0$, in the $F_{{r}t}({r})$ graph
points on the appearance of a returning point or a cusp in the
parametric plot of $\mathcal{L}(\mathcal{F})$, the graph of
$\mathcal{L}(\mathcal{F})$ corresponds then to a multiple--valued
relation. Nevertheless, this multiple--valued property of
$\mathcal{L}(\mathcal{F})$ is not an impediment for the existence of
analytic, in their dependence on the variable $r$,
gravitational--electric ``regular at the center and Maxwell at
infinity'' solutions. This lack of analyticity in the relation
$\mathcal{L}(\mathcal{F})$ is not worse than the infinity at the
origin of the magnetic invariant $\mathcal{F}_m$. Some remarks about
analyticity can be found in \cite{BronnikovPRD2002}.

\section{Magnetic static spherically symmetric
metrics}\label{magnetic}

All SSS gravitational fields coupled to pure magnetic NLE possess a
common field with component $F_{\theta\phi}=h_0 \sin{\theta}$ and a
singular at the origin magnetic field invariant of the form
$2\,\mathcal{F}_m=h_0^2/r^4$; therefore one should strictly call the
Einstein--NLE magnetic solutions singular ones, nevertheless the
associated gravitational field may show a regular behavior of the
curvature invariants, thus one may have solutions with a singular
behavior in the magnetic field invariant but a regular behavior in
the curvature invariants, i.e., semi regular or singular--regular
hybrid. Moreover, any magnetic solution to NLE, is determined by a
single first order differential equation for $Q(r)$ arising from the
$E^t_t$ equation (\ref{EQET1})
\begin{eqnarray}\label{EinsteinL4}&&
\mathcal{L} \left( r \right) =\frac{1}{r^2} +\frac{Q}{r^4}-\frac
{\dot{Q}}{{r}^{3}}-\Lambda, \text{integrating,} \,\nonumber\\&&
Q(r)=Q_K+Q(\mathcal{L}),\,Q_K= r^2-2mr
-\frac{\Lambda}{3}r^4,\,\nonumber\\&& Q(\mathcal{L}):=-r\,\int
r^2\,\mathcal{L} \left( r \right) dr,
\end{eqnarray}
characterized by
\begin{eqnarray}\label{EinsMG}
 &&S(\ref{Sein})=-
\frac{r}{4}\frac{d\mathcal{L}}{dr},\,
\Psi_2=\Psi_2(\ref{curatPsi2}),\nonumber\\&&
\,R=4\mathcal{L}-4\mathcal{L}_{\mathcal{F}}\mathcal{F}_m
+4\Lambda=4\mathcal{L}+r\frac{d\mathcal{L}}{dr}+4\Lambda,
\nonumber\\&& {\mathcal{F}}_m=\,\frac {
 h_0^{2}}{2{r}^{4}}, \,\mathcal{L}_\mathcal{F}=-\frac{d\mathcal{ L}}{dr}\,\frac{r^5}{2h_0^2}.
\end{eqnarray}
The substitution of $\mathcal{L}$ from (\ref{EinsteinL4}) into the
right hand side of $S(\ref{Sein})$ in (\ref{EinsMG}) yields the
identity $S(\ref{Sein})=S(\ref{Sein})$. The regularity of
$\Psi_2\,(\ref{curatPsi2})$, and $R\,( \ref{curvaR})$ for the
magnetic metric requires
\begin{eqnarray}\label{Mag1}
\lim_{{r}\rightarrow 0}\{Q,\dot{Q},\ddot{Q}\}=\{0,0,2\},
\end{eqnarray}
for zero ``mass'' $m$. For the regularity of $S $, and $R(
\ref{EinsMG})$ at $r=0$, one has to establish the regular behavior
of $r{d\mathcal{L}}/{dr}$ and $\mathcal{L}$ at the origin, namely
\begin{eqnarray}\label{Mag2} \lim_{{r}\rightarrow
0}\{\mathcal{L},r\frac{d\mathcal{L}}{dr}\}=\{FQ,FQ'\}.
\end{eqnarray}
Under the fulfilling of the conditions (\ref{Mag1}) and
(\ref{Mag2}), one may consider that a magnetic solution at the
center approaches to the flat or (A)dS spacetimes, although one has
to recall the singularity of the magnetic field invariant
$\mathcal{F}_m$ there. Magnetic SSS solutions to Einstein--NLE
equations, in the $H(P;Q)$--formulation, has been studied previously
by Bronnikov \cite{BronnikovPRD2002}.

\subsection{Linear superposition of SSS magnetic solutions of the
Einstein--NLE equations}\label{superposition}

 Since in the magnetic case, the single Einstein--NLE equation
on $\mathcal{L}$ (\ref{EinsteinL4}) depends linearly on the
structural function $Q(r)$ and its derivatives, the linear
superposition of SSS magnetic solutions $Q_i$ holds too; for each
solution $Q_i$, equipped with its own set of parameters
$p_i=\{{p_{i}}_k,\, i=1,...,n,\,\,k=1,...,s\}$, one determines
$\mathcal{L}_i$ for a common magnetic field
${F}_{\theta\phi}=h_0\sin{\theta}$.
 The linear superposition
of $Q_i(r)$ gives rise to a new enlarged total solution $Q_T=\sum_i
C_i Q_i$ such that $\mathcal{L}_T=\sum_i C_i \,\mathcal{L}_i$,
characterized by the curvature invariants $S_T=\sum_i C_i S_i$,
$\Psi_{2T}=\sum_i C_i \Psi_{2i}$, and $R_T=\sum_i C_i R_i$. Of
course, one may consider that a Lagrangian function $\mathcal{L}(r)$
is given, and one integrates for $Q(r)$; in this way the
superposition of $\mathcal{L}_i$, $\sum_iC_i\mathcal{L}_i$, leads
the superposition $\sum_iC_i Q_i$.

\section{Final remarks}

Summarizing the achievements of the present work devoted to static
spherically symmetric spacetimes with nonlinear electrodynamics
sources, we can mention, among others, the general linear integral
representation of the structural function $Q(r)$ through arbitrary
given electric fields $\mathcal{E}:=q_0 F_{rt}$. Another important
result is the determination of the four Riemann quadratic
invariants, according to the definitions of \cite{McIntoshZakhary97}
and \cite{TorresFayos17}, for a SSS Petrov type $D$ spacetime
coupled to Segre type $[(1,1)(11)]$ NLE, through three (first degree
Riemannian) curvature invariants, namely, the Weyl tensor matrix
eigenvalue $\Psi_2$, the traceless Ricci tensor matrix eigenvalue
$S$, and the curvature scalar $R$. These three invariant functions
depend linearly on the single
 metric function $Q(r)=-r^2g_{tt}$ and its first and second order
 derivatives. Because of their general character, they constitute
 an efficient tool for the full characterization of the
 curvature properties of classes of spacetimes; on the contrary,
 the single Kretschmann quadratic invariant,
 $R_{\alpha\beta\gamma \delta}R^{\alpha\beta\gamma
\delta}$, one of the four Riemannian invariants, which, in the
studied case is the sum of squares of $\Psi_2$,\,$S$, and $\,R $, in
any case, will provide only partial information on this respect. The
regularity conditions, i.e., the requirements for the absence of
singularity of those mentioned curvature invariants at the center
have been established: the Einstein--NLE theory allows for regular
electric SSS solutions with regular curvature invariants at the
origin $\lim_{{r}\rightarrow 0}\{S,\Psi_2,R\}=
\{0,0,(0,4\,\Lambda+4\,\mathcal{L}(0))\}$ if and only if the
following regularity conditions at the center hold:$
\lim_{{r}\rightarrow 0}\{Q,\dot{Q},\ddot{Q}\}=\{0,0,2\}$, and
$\lim_{{r}\rightarrow
0}\{{\mathcal{E}},\dot\mathcal{E},\ddot\mathcal{E}\} $
$=\,\{0,0,0\}$, $\mathcal{E}:=q_0F_{rt}$, see details in
(\ref{RegCondN}); necessary (\ref{NecessaryN}), and sufficient
(\ref{SufficientN}). Hence, regular NLE SSS solutions approach to
the conformally flat dS--AdS or flat spacetimes regular at the
center. From the weak--dominant energy conditions \ref{WECDC}, the
inequalities that have to be fulfilled for physically reasonable NLE
fields are established; among them the Lagrangian function--energy
density relation:
$\mu(r)=\mathcal{L}(r)-2\mathcal{F}\mathcal{L}_\mathcal{F}$, which
at the center yields $\mu(0)=\mathcal{L}(0)\geq 0$, for acceptable
physical solutions.

Next, the linear superposition properties of electrically charged
solutions have been established, \ref{LinearQ}: for static
spherically symmetric electric metrics coupled to NLE and a
cosmological constant, any linear superposition of metric functions
$Q_i(r):=Q(\mathcal{E}_i)$ yields to the linear superpositions of
Lagrangian functions $\mathcal{L}_i=\mathcal{L}(Q_i)$ and the
corresponding electromagnetic field functions
$\mathcal{E}_i;={q_0}_i{F_{yt}}_{i}$, no $\sum$ in $i$, which, in
turn, will be solutions of the Einstein--electrodynamics field
equations too. This property can be extended to the pure magnetic
NLE SSS metrics, having in mind that the magnetic field
$F_{\theta\,\phi}=h_0\sin{\theta}$ is sheared by all magnetic
spacetimes. \\ Moreover, in the electrically charged case, since the
invariant functions $\Psi_2$, $r^2S=-q_0F_{rt}/2$, and $\mathcal{L}$
are described by Euler equations for the metric function $Q(r)$,
then, in general, one can determine NLE SSS solutions with
preestablished properties of those functions, for instance, the
metric for a particular Weyl curvature structure $\Psi_2$, see
\ref{EulerCurInv}\,(\ref{QspecPSI}).

Although black holes have been discovered and described
theoretically more than half a century ago, it has been only until
recently that they have been experimentally found indirectly on
September 14, 2015. The Laser Interferometer Gravitational--Wave
Observatory (LIGO) detected the gravitational wave GW150914,(LIGO
Virgo collaboration)~\cite{Abbott}, emitted by a binary system of
rotating black holes many thousand light years ago; this first wave
detection showed indirectly the existence of spinning black holes.
New investigations in experimental black hole physics have been
undertaken since that discovery. The list of publications since that
date is quite large, in this respect, we cite some
articles  published recently in Physical Review Letters:\\
Clovecko et al. \cite{Clovecko2019} used `` the spin--precession
waves propagating on the background of the spin super--currents
between two Bose--Einstein condensates of magnons... as an
experimental tool simulating the properties of the
black--and white--hole horizons.'' \\
Hughes et al. \cite{Hughes2019} studied ``...the coalescence of two
black holes which generates gravitational waves that carry detailed
information about the properties of those black holes and their
binary configuration.''\\
Nair et al. \cite{Nair2019}  presented ``...a study of whether the
gravitational--wave events detected so far by the LIGO--Virgo
scientific collaborations can be used to probe higher-curvature
corrections to general relativity.''\\
Yang et al. \cite{Yang2019} showed ``...that if migration traps
develop in the accretion disks of active galactic nuclei (AGNs) and
promote the mergers of their captive black holes, the majority of
black holes within disks will undergo hierarchical mergers—-with one
of
the black holes being the remnant of a previous merger.''\\
Pook et al. \cite{Pook2019} found ``...strong numerical evidence for
a new phenomenon in a binary black hole spacetime, namely, the
merger of marginally outer trapped surfaces (MOTSs). By simulating
the head-on collision of two nonspinning unequal mass black holes,
we observe that the MOTS associated with the final black hole merges
with the two initially disjoint surfaces corresponding to the two
initial black holes.''\\
Baumgarte et al. \cite{Baumgarte2019} ``...numerically investigated
the threshold of black-hole formation in the gravitational collapse
of
electromagnetic waves in axisymmetry.''\\
Coates et al. \cite{Coates2019} studied `` ...black hole area
quantization in the context of gravitational wave physics.''\\
Abbott et al. (LIGO Scientific Collaboration and the Virgo
Collaboration) \cite{LIGOV2019}
 presented ``...a search for subsolar mass ultra--compact objects in data
obtained during Advanced LIGO's second observing run. In contrast to
a previous search of Advanced LIGO data from the first observing
run, this search includes the effects of component spin on the
gravitational waveform.''

These reports show the wide spectrum of research themes in the area
of detecting gravitational waves to confirm the existence of black
holes in the Universe.

ACKNOWLEDGMENTS. GGC acknowledges the support of Consejo Nacional de
Ciencia y Tecnolog\'{\i}a (CONACYT) through a doctoral fellowship.

\end{document}